\documentclass[twoside,11pt]{article}

\usepackage{jmlr2e}

\usepackage{url}
\usepackage{hyperref}
\usepackage[normalem]{ulem} 
\usepackage{verbatim}
\usepackage{enumerate,mdwlist}
\usepackage[titletoc]{appendix}
\usepackage{upgreek}
\usepackage{amsfonts}
\usepackage{physics}
\usepackage{color}
\usepackage{booktabs}
\usepackage{xcolor}
\usepackage{algorithm2e}

\RestyleAlgo{ruled}

\newcommand{\p}{\boldsymbol{\Pi}}
\newcommand{\x}{\boldsymbol{x}}
\newcommand{\U}{\boldsymbol{u}}

\usepackage{lastpage}
\jmlrheading{24}{2023}{1-\pageref{LastPage}}{12/22}{8/23}{22-1450}{Jakob Robnik, G. Bruno De Luca, Eva Silverstein and Uro\v{s} Seljak}

\ShortHeadings{Microcanonical Hamiltonian Monte Carlo}{Robnik, De Luca, Silverstein and Seljak}

\begin{document}

\title{Microcanonical Hamiltonian 
Monte Carlo}

\author{\name Jakob Robnik \email jakob\_robnik@berkeley.edu \\
      \addr Physics Department\\
      University of California at Berkeley\\
      Berkeley, CA 94720, USA
      \AND
      \name G. Bruno De Luca \email gbdeluca@stanford.edu \\
      \addr Stanford Institute for Theoretical Physics\\
      Stanford University\\
      Stanford, CA 94306, USA
      \AND
      \name Eva Silverstein \email evas@stanford.edu \\
      \addr Stanford Institute for Theoretical Physics\\
      Stanford University\\
      Stanford, CA 94306, USA
      \AND
      \name Uro\v{s} Seljak \email useljak@berkeley.edu \\
      \addr Physics Department\\
      University of California at Berkeley\\ and Lawrence Berkeley 
      National Laboratory\\
      Berkeley, CA 94720, USA}
       

\editor{Ryan Adams}
 
\maketitle

\begin{abstract}
We develop Microcanonical Hamiltonian Monte Carlo (MCHMC), a class of models that follow fixed energy Hamiltonian dynamics, in contrast to Hamiltonian Monte Carlo (HMC), which follows canonical distribution with different energy levels. MCHMC tunes the Hamiltonian function such that the marginal of the uniform distribution on the constant-energy-surface over the momentum variables gives the desired target distribution. 
We show that MCHMC requires occasional energy-conserving billiard-like momentum bounces for ergodicity, analogous to momentum resampling in HMC. We generalize the concept of bounces to a continuous version with partial direction preserving bounces at every step, which gives energy-conserving underdamped Langevin-like dynamics with non-Gaussian noise (MCLMC). 
MCHMC and MCLMC exhibit favorable scalings with condition number and dimensionality. We develop an efficient hyperparameter tuning scheme that achieves high performance and consistently outperforms NUTS HMC on several standard benchmark problems, in some cases by orders of magnitude. 
\end{abstract}


\section{Introduction}

Sampling is an important element of various scientific disciplines, ranging from quantum chromodynamics and statistical physics to economics and Bayesian inference. The need for samplers arises from the need to compute expectation values of the functions $\mathcal{O}(\x)$ of the high dimensional parameters $\x$, given the parameter probability distribution $p(\x) = e^{- \mathcal{L}(\x)} / Z$. Typically, we have access to $\mathcal{L}(\x)$, but 
not to the normalization $Z$. 
Computing the expectation value integral and the normalization $Z$ with a brute force 
integration is prohibitively expensive. An alternative is to construct a sampler - an algorithm which generates a stream of vectors $\{ \x_n \}_{n = 1}^{N}$, distributed according to the target distribution $p(\x)$. Taking the expectation value is then a simple matter of summing $\langle \mathcal{O} \rangle \approx \frac{1}{N} \sum_{n = 1}^{N} \mathcal{O}(\x_n)$. This is an easier task than performing the integral because the computational resources are not wasted in regions where the probability mass $p(\x)$ is low.
\par
A general class of sampling models
is Monte Carlo Markov Chain (MCMC), 
which uses detailed balance for 
transitions between the chain 
elements. 
A gold standard for MCMC sampling with 
available gradient $\nabla \mathcal{L}(\x)$ is Hamiltonian Monte Carlo \citep{HMCDuane, HMCneal,conceptualHMC}. It promotes the original $d$-dimensional $\x$-space to a $2 d$-dimensional phase-space, with the addition of the canonical momentum $ \p$. It translates the task of sampling in the $\x$-space with the target $p(\x)$ to the task of sampling on the phase-space with the canonical ensemble target $p(\x, \p) \propto \exp{- H(\x, \p)}$. The Hamiltonian function is tuned in such a way that the marginal of the phase-space target over the momentum coordinates gives the original target. The most popular choice is $H(\x , \p) = \frac{1}{2} \vert \p \vert^2 +\mathcal{L}(\x)$. Sampling on the phase space is convenient because the phase space can be split into the surface levels of the Hamiltonian function, and each surface can be efficiently explored by simulating the Hamiltonian dynamics, which preserves the Hamiltonian function, that is, the energy. The transitions between the energy surfaces are achieved by the occasional momentum re-sampling according to its marginal distribution - a Gaussian. However, the convergence can be slow if mixing becomes inefficient \citep{conceptualHMC}, in which case the 
samples are highly correlated. A related
dynamics with similarly good sampling properties is that of underdamped or 
overdamped Langevin MC \citep{MolecularDynamics}.

\begin{figure}
    \makebox[\textwidth][c]{\includegraphics[scale = 0.55]{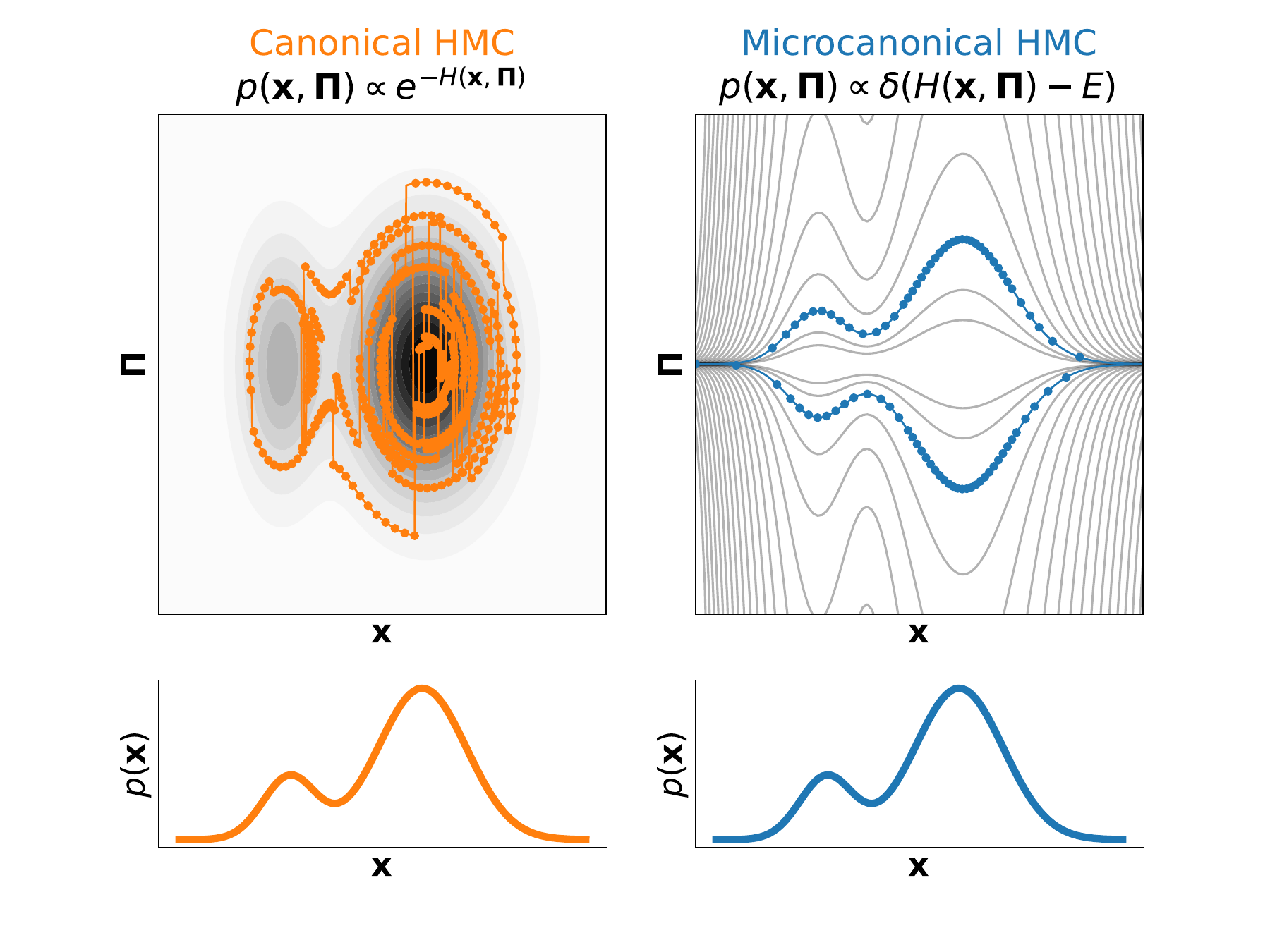}}
    \caption{We illustrate the difference between the canonical HMC (left) and microcanonical HMC (right).
    The canonical HMC target on the phase space is $p(\x, \p) \propto e^{-H(\x, \p)}$ (gray scale). The Hamiltonian is tuned such that the marginal over the momentum variables (bottom) gives the desired target distribution $p(\x)$. The dynamics (orange points) is a combination of the Hamiltonian evolution and momentum resampling. Typically, the points just before the resampling are collected to form samples, possibly with the metropolis adjustment step inserted. In multinomial HMC \citep{conceptualHMC}, the samples are drawn from all points, each with the weight $\propto e^{-H}$.
    The Microcanonical HMC target on the phase space is a single energy level $p(\x, \p) \propto \delta(H(\x, \p) - E)$. Energy levels are shown with gray lines. Again, the Hamiltonian is tuned such that the marginal $p(\x)$ is the desired target. The dynamics (blue points) is a combination of the Hamiltonian evolution and momentum bounces. All the integration points are used as samples.
    } 
    \label{fig: illustration}
\end{figure}

In this paper, we first introduce a class of 
models we call Microcanonical Hamiltonian Monte Carlo (MCHMC), which do not  resample energy. Instead, we tune the Hamiltonian function such that the marginal of the uniform distribution on the constant-energy-surface (known as the microcanonical ensemble in physics) over the momentum variables gives the desired target distribution $p(\x)$. 
The distinction between the canonical HMC and microcanonical HMC is illustrated in Figure \ref{fig: illustration}.
A specific deterministic model from this 
class has recently been introduced as 
Energy Sampling Hamiltonian (ESH) by \cite{ESH}. Here we 
show that there are 
infinitely many Hamiltonians in this class. While
in the main text we emphasize
the Hamiltonian nature of the MCHMC
sampler, in Appendix \ref{sec: geodesic} 
we explore the Riemannian geometry view,
which covers an even broader class of 
models and provides another 
generalization path.

Sampling from a target distribution 
can be achieved using either
stochastic or deterministic 
methods, but it has been argued
that deterministic methods 
have less noise and 
converge faster than 
stochastic methods \citep{DLMC}. ESH 
is a
deterministic algorithm 
\citep{ESH}, and would potentially carry similar benefits. However, 
we show that ESH is in general not ergodic and does not 
converge to the target distribution. In \cite{ESH} this was partially addressed by 
having many independent chains, 
such that each chain provides only 
one sample at the end of its run. 
We show that this does not resolve the 
problem with ergodicity, and the
ESH algorithm does not converge 
to the true posterior even when the chain 
length is long, because random 
initialization 
combined with deterministic ESH dynamics does 
not guarantee convergence to the 
true target in the 
limit of large number of initial 
starting points. 

    In this paper we
    propose two different solutions: in the first we complement the dynamics with occasional  random momentum bounces, which encourage rapid exploration of the energy surface,  as was recently used for optimization with microcanonical Hamiltonians in \cite{BIoptimization}. These bounces
    play a similar role as the momentum 
    resampling in HMC, except that 
    the energy is conserved during the 
    bounce. 
    In the second
    solution we apply
    bounces at every step, but with
    partial preservation of the 
    momentum direction, which  
    we call Microcanonical 
    Langevin-like Monte Carlo (MCLMC). It can be 
    viewed as a Langevin-like dynamics 
    with non-Gaussian noise. 
    With the random 
    bounces, the MCHMC and MCLMC
    algorithms we propose 
    are no longer 
    deterministic, in contrast to ESH. However, 
    they can still exhibit less
    noise than other MCMC
    algorithms, and potentially 
    converge faster to the 
    target distribution. 

    Tuning MCMC 
    samplers is often 
    an expensive and poorly understood 
    procedure (see e.g. \citet{NUTS}), and 
    methods where the tuning can be 
    reduced or 
    avoided completely have a distinct advantage. 
    We develop a fast tuning algorithm for bounce frequency (bounce strength for MCLMC) and the integration step-size. 
    We test our method on various benchmark problems in Section \ref{sec: results}. 
The code with a tutorial is publicly available\footnote{\url{https://blackjax-devs.github.io/sampling-book/algorithms/mclmc/}}.

\section{Method} \label{sec: method}

The Hamiltonian equations describe the time evolution of the generalized position $\x(t)$ and conjugate momenta $\p(t)$ of the classical physical systems:
\begin{equation} \label{eq: hamilton eqs}
    \dot{\x} = \frac{\partial H(\x, \p)}{\partial \p} \qquad \dot{\p} = -\frac{\partial H(\x, \p)}{\partial \x},
\end{equation}
where $H(\x, \p)$ is called the Hamiltonian function or the energy. The system evolving under the Hamiltonian equations remains forever bound to the constant energy surface:
\begin{equation}
    \frac{d}{dt} H(\x(t), \p(t)) = \frac{\partial H(\x, \p)}{\partial \x} \dot{\x} + \frac{\partial H(\x, \p)}{\partial \p} \dot{\p} = 0.
\end{equation}
Ergodicity is an additional assumption that the dynamics visits every part of the energy surface uniformly. Then, for an arbitrary observable $f(\x, \p)$, the ensemble average equals the time average
\begin{equation} \label{eq: time ensemble average}
    \int_{\mathbb{R}^{2 d}} f(\x, \p) p_{E}(\x, \p) d \x d\p = \lim_{T \xrightarrow[]{} \infty} \frac{1}{T} \int_0^T f(\x(t), \p(t)) dt,
\end{equation}
where $p_E(\x, \p)$ is the uniform distribution on the constant energy surface $p_E(\x, \p) \propto \delta(H(\x, \p) - E)$, also called the microcanonical ensemble.
Ergodicity makes the ensemble expectation values practical to compute because one can simulate the Hamiltonian dynamics and take the time average over the trajectory.

\subsection{Tuning the Hamiltonian}\label{sec: tunHam}
The idea of MCHMC is to tune the Hamiltonian function in a way that the microcanonical ensemble, marginalized over the latent momentum variables, gives the desired target distribution:
\begin{equation}\label{eq: marginal condition}
    p(\x) \propto \int_{\mathbb{R}^d} \delta(H(\x, \p) - E) d \p = \frac{\Omega_{d-1} \Pi^{d-1}}{\vert \partial_{\p} H\vert},
\end{equation}
where $\Pi = \vert \p \vert$. In the last step, we assumed that the Hamiltonian does not depend on the direction of the momentum, only on its magnitude. The delta function condition then fixes the magnitude of the momentum at each $\x$ and the angular part of the integral gives a volume of the $d-1$ dimensional unit sphere, $\Omega_{d-1}$.

We see that the target distribution is a result of two effects. The first is the number of momentum states the energy surface has at a given $\x$; this is the factor $\Omega_{d-1} \Pi^{d-1}$. The second is the number of samples the trajectory will generate when passing through a small neighborhood of $\x$. This is proportional to the time it will spend there, so inversely proportional to the magnitude of the velocity, that is $1 / \vert \partial_{\Pi} H \vert = 1 / \vert \dot{\x}\vert$, which is the second factor in Equation \eqref{eq: marginal condition}.

The proposed MCHMC is a class of 
models, as 
we have considerable freedom in choosing the Hamiltonian on the phase space. 
Without striving for completeness, we consider several physics-inspired options here.  
One important class are separable 
Hamiltonians,
\begin{equation}
    H(\x,\p)=E=T(\p)+V(\x), 
    \label{separable}
\end{equation}
where $T(\p)$ is kinetic term and 
$V(\x)$ is potential term. We will 
make further simplification by assuming $T$ is a function of the momentum magnitude $\Pi$ only and will take a class of functions labeled by a continuous index $q$: 
\begin{equation}
    T(\p) = 
    \begin{cases}
    \log \Pi & q = 0\\
    \frac{\Pi^q}{q} & q \neq 0
    \end{cases},\label{eq:qcases}
\end{equation}
such that the speed $\vert \dot{\x} \vert = T'(\Pi)= \Pi^{q-1}$ is a power law of the momentum.
In this case, the marginal condition \eqref{eq: marginal condition} determines the potential function $V(\x)$ in terms of the target density:
\begin{equation} \label{eq: V}
    V(\x) = E - T \bigg( \exp{-\frac{\mathcal{L}(\x)}{d - q}} \bigg) = 
    E + \begin{cases}
    d^{-1} \mathcal{L}(\x) &  q = 0\\
    - q^{-1} e^{- q \mathcal{L}(\x) / (d- q)} & q \neq 0
    \end{cases},
\end{equation}
as can be seen by solving Equation \eqref{eq: marginal condition} for $\Pi$ and applying the function $T$ on both sides.

\subsubsection{$q=0$: variable mass Hamiltonian} \label{sec: pdm}

Up to a time rescaling and an irrelevant energy shift, the $q = 0$ Hamiltonian is equivalent to the one proposed in \cite{ESH}:
\begin{equation} \label{eq: ESH}
    H(\x, \p) = \frac{d}{2} \log \frac{\p^2}{d} + \mathcal{L}(\x).
\end{equation}

Interestingly, this Hamiltonian has the same dynamics as the kinetic Hamiltonian with the position dependent mass:
\begin{equation} \label{eq: H1}
    \widetilde{H}(\x, \p) = \frac{\p^2}{2 m(\x)},
\end{equation}
where the mass is given by the target density:
\begin{equation}\label{eq: tuning1}
    m(\x) = e^{- 2 \mathcal{L}(\x)/d}. 
\end{equation}
This is because the Hamiltonians are related by the transformation $\exp{ 2 H / d} = 2 \widetilde{H} / d$.
The mass is a monotonically increasing function of the target density, making the particle move more slowly in the high-density regions.

The separable form of the Hamiltonian \eqref{eq: ESH} enables development of efficient integrators. Following \citet{ESH}, we first write the Hamiltonian equations as \eqref{eq: hamilton eqs}
\begin{equation} \label{eq: EOM}
    \dot{x} = \frac{\p}{\vert \p \vert} \frac{1}{w} \qquad
    \dot{\p} = - \nabla \mathcal{L}(\x),
\end{equation}
where $w = \vert \p \vert / d$. Directly integrating the Hamiltonian equations is suboptimal because the step-size must be kept small to accurately capture the U-turn that the particle makes in low density regions where the speed is high. An adaptive algorithm where all the steps make the same length along the trajectory is desired once we reach the typical set. We introduce the natural parameter $d s = d t / w$ and denote $\frac{d}{d s}$ by a dot from now on. The Hamiltonian equations become
\begin{align} \label{eq: rescaled EOM}
    \dot{\x} &= \U\\ \nonumber
    \dot{\U} &= - d^{-1}({\bf I} - \U \U^T) \nabla \mathcal{L}(\x)\\ \nonumber
    \frac{\dot{w}}{w} &= - d^{-1}\U \cdot \nabla \mathcal{L}(\x),
\end{align}
where $\U = \frac{\p}{\vert \p \vert}$ is the direction of the momentum. The first equation establishes the natural parameter as the length along the trajectory. The second Equation establishes the dynamics as a rotation of the momentum orientation towards the direction of the target log-density gradient. The larger the gradient, the faster is this rotation. The magnitude of $\U$ is preserved because the derivative $\dot{\U}$ is perpendicular to $\U$. Note that the dynamics is independent of the magnitude of the momentum and therefore independent of the energy.

Only the weights $w$ depend on the energy, but they do not affect the 
dynamics. Instead of using 
Equation \eqref{eq: rescaled EOM} we can evaluate the 
weight from Equation \eqref{eq: ESH},
\begin{equation}
    w=\exp[(E-\mathcal{L})/d],
    \label{ECW}
\end{equation}
where the MCHMC dynamics is on a constant Hamiltonian surface $H(\x, \p)=E$. The integration steps 
that are evaluated with a constant step size in natural parameter $s$
must be reweighted by $w$ to be 
uniform in time $t$. However, 
since the weights are renormalized
to add up to unity the dependence on $E$ drops out and $q=0$ MCHMC is
independent of the choice of overall energy $E$. Note that in high dimensions 
for near Gaussian targets the spread of the typical set is 
$\Delta \mathcal{L} =O(d^{1/2})$, 
so the spread of the 
weights is $\Delta \ln w= O(d^{-1/2})$, i.e. the weights become nearly constant
for high $d$. In this paper, we will 
use the last of Equation \eqref{eq: rescaled EOM} to 
evaluate the momentum amplitude and the 
energy through Equation \eqref{eq: ESH}. 
This allows us to monitor the error in 
energy conservation, which we will use 
to set the step size of our 
integrator. We have verified that the 
required error in energy needs to 
be sufficiently small such that 
there is not much difference in terms 
of which expression we use for the 
weights, but the energy conservation 
weight of Equation \eqref{ECW} is more stable
for larger stepsize $\epsilon$. 

An integrator is an update rule $\Phi_{\epsilon}$ which pushes the quantities $\x(n \epsilon) = \x_n$, $\U(n \epsilon) = \U_n$ and $w(n \epsilon) = w_n$ by an amount $\epsilon$ forward in $s$.
The simplest second order integrator is the leapfrog integrator. The velocity-leapfrog integrator first updates the momentum by half step, then the position by a full step and then again the momentum by the remaining half step:
\begin{equation} \label{eq: leapfrog}
    (\x_{n+1},\, \U_{n+1}, w_{n+1}) = \Phi_{\epsilon}(\x_n, \, \U_n, \, w_n) = (\Phi_{\epsilon / 2}^{V} \circ \Phi_{\epsilon}^{T} \circ \Phi_{\epsilon / 2}^{V}) (\x_n, \, \U_n, \, w_n).
\end{equation}
It requires only one target density gradient evaluation per step. This is the integrator typically used in HMC and was also adopted in \cite{ESH}, where the maps $\Phi_{\epsilon}^T$ and $\Phi_{\epsilon}^V$ were derived. The position updating map is
\begin{equation} \label{eq: leapfrog position}
    \Phi^T_{\epsilon}(\x, \U, w) = (\x + \epsilon \U, \U, w),
\end{equation}
and the momentum updating map is
\begin{equation} \label{eq: leapfrog momentum}
    \Phi_{\epsilon}^V(\x, \U, w) = \bigg( \x, \,
    \frac{\U + (\sinh{\delta}+ \boldsymbol{e} \cdot \U (\cosh \delta -1)) \boldsymbol{e} }{\cosh{\delta} + \boldsymbol{e} \cdot \U \sinh{\delta}}, \,
    w \,(\cosh \delta + \boldsymbol{e} \cdot \U \sinh \delta) \bigg), 
\end{equation}f
where $\delta = \epsilon \vert \nabla \mathcal{L}(\x) \vert / d$ and $ \boldsymbol{e} = - \nabla \mathcal{L}(\x) / \vert \nabla \mathcal{L}(\x) \vert$.
Note that the leapfrog integrator of Equations \eqref{eq: rescaled EOM} is not symplectic, while the leapfrog integrator of Equations \eqref{eq: EOM} would be.

\cite{MinimalNorm} introduced the Minimal Norm (MN) integrator, which is also second order, but additionally designed to minimize the coefficients of the third order residuals. It requires two gradient evaluations per step, but is expected to allow for $\sqrt{10.9}$ larger steps, so we expect an efficiency improvement of $65 \%$ \citep{MinimalNormQCD}. It is composed of five sub-steps:
\begin{equation} \label{eq:MN}
    \Phi_{\epsilon} = \Phi_{\epsilon \lambda}^{V} \circ \Phi_{\epsilon/2}^{T}\circ \Phi_{\epsilon (1-2\lambda)}^{V} \circ \Phi_{\epsilon/2}^{T} \circ \Phi_{\epsilon \lambda}^{V},
\end{equation}
where $\lambda = 0.19318...$ \citep{MinimalNormQCD}. 
Table \ref{table} shows that the MN integrator performs better than leapfrog on a majority of benchmark problems.


Initialization can be chosen by the user. Here 
we will randomly draw the initial position $\x_0$ from the prior and draw $\boldsymbol{u}_0$ from an isotropic distribution. The initial weight $w_0 = 1$. Applying the update rule \eqref{eq: leapfrog} several times gives us an approximation to the trajectory at the discrete time steps $\x_n = \x(n \epsilon)$ and the associated weights $w_n$. Under the ergodic hypothesis \eqref{eq: time ensemble average}, the expectation values of interest are then computed as 
\begin{equation}
    \langle \mathcal{O} \rangle = \frac{\sum_{n = 1}^{N} \mathcal{O}(\x_n) w_n}{\sum_{n = 1}^{N} w_n}.
\end{equation}
If the memory cost is important, we do not need to store the samples, but can update the expectation values with a Kalman filter. Starting with $W_0 = w_0$ and $\mathcal{O}_0 = \mathcal{O}(\x_0)$ we update after each step
\begin{equation}
    W_{n+1} = W_n + w_{n+1} \qquad \mathcal{O}_{n+1} = \frac{W_n}{W_{n+1}}\mathcal{O}_{n} +  \frac{w_n}{W_{n+1}}\mathcal{O}(\x_{n+1})\,,
\end{equation}
and output $\langle \mathcal{O} \rangle = \mathcal{O}_N$. Note that computing the marginal histograms also falls under this formalism by taking $\mathcal{O}$ to be an indicator function of the bin.

In our examples below, Hamiltonian evolution can reach $\vert \x \vert \to \infty$ in finite time $t$, necessitating a boundary condition there to complete the specification of the system.  For targets with a typical set consisting of one connected component, and with the rescaled integrator just described, we do not find this to be of practical importance. But contributions to the dynamics from boundary reflections may be of interest in the general case and can contribute to ergodicity in a complementary way to the bounces we use here (see e.g. \cite{PhysRevLett.77.2941}).

\subsubsection{$q=2$: standard kinetic energy Hamiltonian} \label{sec: TV}

The standard canonical form of the kinetic energy is with $q = 2$:
\begin{equation}\label{eq: H3}
    H(\x , \p) = \frac{\p^2}{2} + V(\x) .
\end{equation}

The potential energy is then given by Equation \eqref{eq: V}:
\begin{equation}\label{eq: tuning3}
    V(\x) - E = -\frac{1}{2} e^{- 2 \mathcal{L}(\x) / (d-2)}.
\end{equation}
This can be contrasted to the standard HMC, where $V(\x) = \mathcal{L}(\x)$. Here, as in the standard HMC, the particle moves faster when the density is high, and the method has to make up for this by passing through the high-density regions many times (with different momenta). In contrast, the variable mass 
method moves slower in regions with 
high density, which suggests 
that it will require fewer orbits 
to converge to the target. 

The dynamics is that of a particle in a potential:
\begin{equation}
    \dot{\x} = \p \qquad \dot{\p} = - \nabla V(\x) = - \nabla \mathcal{L}(\x) \, \frac{e^{- 2 \mathcal{L}(\x) / (d-2)}}{d-2}.
    \label{eq:standard}
\end{equation}

Many efficient integrators can be used to solve the Hamiltonian of this type \citep{GeometricNumericalIntegration, MolecularDynamics, SimulatingHamiltonianDynamics}. We tested symplectic Euler, leapfrog, fourth-order Runge-Kutta, and Yoshida's steps and found the best performance with Yoshida's steps. We draw an initial condition $\x_0$ from the prior and determine the initial momentum such that the total energy is zero. Its direction is chosen randomly.

Note that the potential \eqref{eq: tuning3} is ill-defined if $d = 2$. This is not surprising, as we cannot tune the momentum states by changing the potential since the density of the momentum states is energy-independent in two dimensions, as is well known. In this case one could introduce a nuisance parameter $z$, such that the marginal over $z$ is the desired target, for example  $p(x, y, z) = p(x, y) \mathcal{N}(z, 0, 1)$.

\subsubsection{Non-separable relativistic Hamiltonian} \label{sec: relativistic}
An example of a non-separable Hamiltonian is that of a relativistic particle with a variable-speed-of-light $c$:
\begin{equation} \label{eq: H2}
    H(\x, \p) = \sqrt{c^4(\x) + c^2(\x) \p^2} .
\end{equation}

The marginal condition \eqref{eq: marginal condition} gives
\begin{equation}\label{eq: tuning2}
    c(\x)^2 = E \, g^{-1}\left( p(\x)^{-2/d} E^{1 - d/2} \right),
\end{equation}
where $g(m) \equiv m (1 - m^2)^{\frac{2}{d} -1}$ is the dimensionless function we have to invert. 
The limiting speed $c(\x)$ is low in the high-density regions, forcing the particle to move slowly. This property was exploited for optimization in \citet{BIoptimization}. 
A majority of samples are collected in the regime where $m = c(\x)^2 / E \ll 1$ and $g(m) \approx m$. There, the relativistic Hamiltonian has the same behavior as the variable mass Hamiltonian. On the other hand, in the low density region, the relativistic dynamics becomes equivalent to the dynamics of the separable standard kinetic energy Hamiltonian.

Due to the difficulty of constructing an efficient integrator for the relativistic Hamiltonian and its equivalence with the variable mass Hamiltonian in the regime where a majority of samples are produced, we will not analyze the relativistic Hamiltonian in this work. The general purpose symplectic integrators of the non-separable Hamiltonians are available \citep{SymmetricProjection}, but they are less efficient than their separable cousins.

Nevertheless, the relativistic Hamiltonian might be of interest whenever the transition through the low-density region is important. Two prominent examples are sampling from the multimodal posteriors with widely-separated modes and the burn-in stage of sampling. Note that the burn-in samples are not discarded for MCHMC, because the particle's speed is accordingly high in the low density regions.

\subsection{A geometric picture}
While in this work we mostly focus on the Hamiltonian point of view, sampling by Hamiltonian evolution on a fixed energy surface can also be understood geometrically. Specifically, we can ask if it is possible to define a notion of distance in configuration space such that, following \emph{geodesics} with respect to this distance, we visit points distributed according to our target distribution $p(\x)$. The answer is affirmative, and in the framework of Riemannian geometry such a distance can be induced by a metric $g(\x)$, which is not uniquely fixed by these requirements. This freedom is analogous to the freedom in the choice of the Hamiltonian. The idea of mapping the Hamiltonian evolution to geodesic motion on an appropriately defined geometry was already introduced by Jacobi (see e.g. the review in \citet{pettini2007geometry}) and later extended in different contexts. In App.~\ref{sec: geodesic} we review some basic facts of Riemannian geometry and its links with Hamiltonian evolution, deriving the explicit map from the Hamiltonians discussed in \S \ref{sec: tunHam} to the corresponding \emph{Jacobi metrics} $g(\x)$. Specifically, we show (Prop.~\ref{prop: geo}) how microcanical Hamiltonian sampling with isotropic kinetic term (i.e.~only depending on $|\p|$) is equivalent to geodesic motion on a conformally-flat configuration space.

This geometric picture provides a complementary point of view to develop intuition about the dynamics. For example, for the $q=0$ variable-mass Hamiltonian the corresponding Jacobi metric is determined by the requirement that the volume density is proportional to $p(\x)$. Thus the effect of $g(\x)$ is to distort the local geometry such that most of the volume is concentrated where the target is larger, and the exploration naturally spends more time in those regions, collecting more samples there. As shown in App.~\ref{sec: geodesic}, in the more general case there is a combined effect of local volume distortion with a distortion of the geodesics ``time". In addition, curvature properties of $g(\x)$ control the spreading of geodesics and thus the amount of intrinsic exploration, before any addition of dispersing elements. An interesting open question is how to exploit more the geometric picture to construct larger new classes of samplers with favorable properties.

\subsection{Momentum decoherence} \label{sec: decoherence}

It is obvious that the ergodicity will not hold if the target has some symmetries (for example, the standard Gaussian with the rotational SO($d$) symmetry).  Noether's theorem \citep{noether} then guarantees the existence of conserved quantities which limit the movement of the sampler to some subset of the energy surface. 
However, even if there are no symmetries, 
ergodicity is not guaranteed. 
In \cite{ESH} 
the proposed strategy is to run many independent chains starting from some random 
initialization from the prior. We will 
show this strategy does not ensure 
ergodicity. 
Here we propose a different strategy, 
using momentum decoherence at fixed energy. A similar strategy is used 
in the context of HMC, where momentum 
is fully resampled and energy 
changes. Here, we will
explore momentum resamplings that 
conserve energy.
Additional strategies for enhancing chaotic behavior might include boundary reflections as mentioned above (cf \cite{PhysRevLett.77.2941}), and pursuing the connection between Hamiltonian dynamics and geodesic motion on curved geometries described in Appendix \ref{sec: geodesic}.

\begin{figure}
    \makebox[\textwidth][c]{\includegraphics[scale = 0.32]{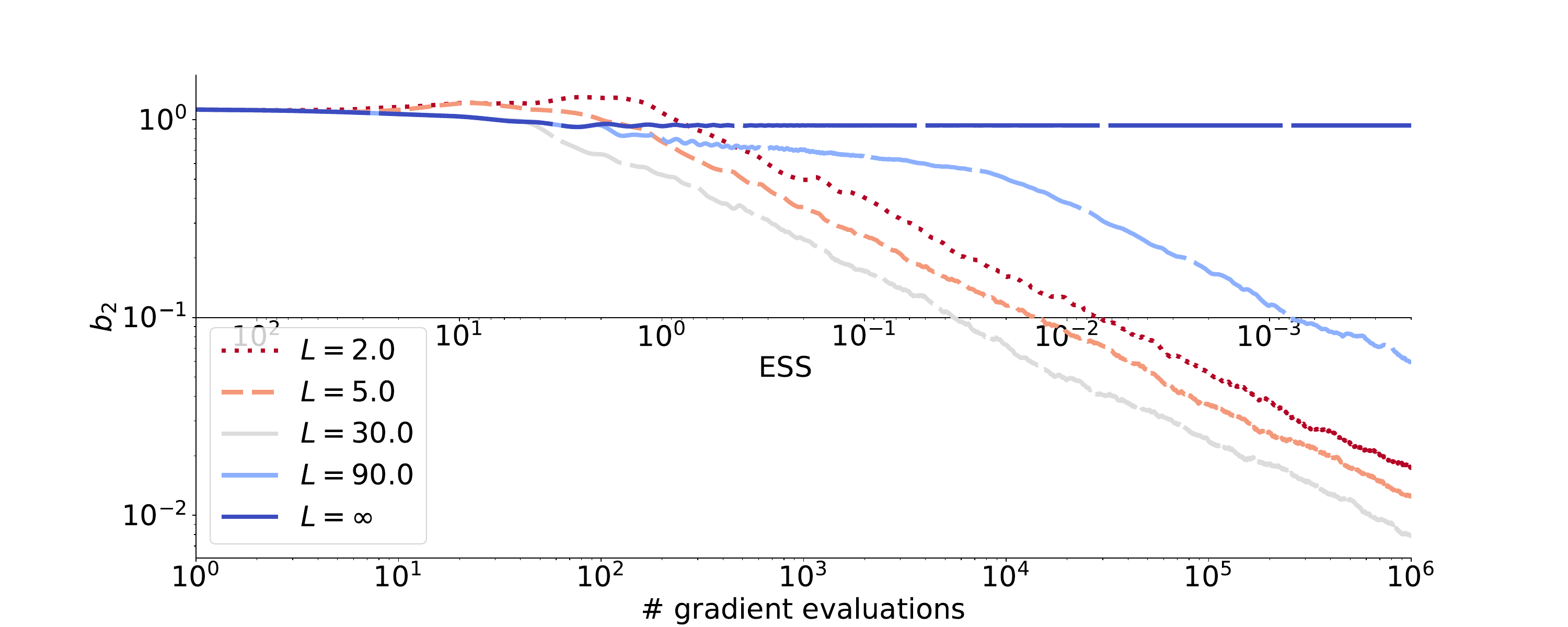}}
    \caption{We simulate the Hamiltonian dynamics of the variable mass Hamiltonian Equation \eqref{eq: H1} to sample from the standard 200-dimensional Gaussian target. A bounce occurs when we move by distance $L$. The square root variance of the second moment $b_2$ of Equation \eqref{eq:b} is shown as a function of the number of the target gradient evaluations. The convergence is slow if the frequency of bounces $L^{-1}$ is too low, because the dynamics is not mixing enough, and
    a single chain never converges ($L = \infty$).
    Because of the random walk behavior, the convergence is also slower if the bounce frequency is too large. ESS is defined as 200 divided by the number of gradient evaluations when the sampler crosses the $b_2 = 0.1$ line. Note that we did not optimize for step size $\epsilon$, for which we use $\epsilon=1$, so ESS is lower than for optimal step size, and $L$ equals the number of steps taken between the bounces.} 
    \label{fig: bounces}
\end{figure}

\subsubsection{Bounces} \label{sec: bounces}

\begin{algorithm}
\caption{MCHMC $q=0$ algorithm.}\label{alg: bounces}
\KwData{
$\text{initial condition } \x_0 \in \mathbb{R}^d$, \\
$\text{number of samples } N > 0$,
$\text{step size } \epsilon > 0$, \\
$\text{steps between the bounces } K = L/\epsilon \in \mathbb{N} $.
}
\KwResult{$\text{samples} \{ \x_n \}_{n = 1}^N$, $\text{weights} \{ w_n \}_{n = 1}^N$}
$w_0 \gets 1$\;

\For{$n\gets0$ \KwTo $N$}{
    \If{$n \text{ is divisible by } K$}{
        $\U_n \gets \text{isotropic random unit vector}$;
    }
    $\x_{n+1}, \, \U_{n+1}, \, w_{n+1} \gets \Phi_{\epsilon}(\x_n, \U_n, w_n)$, see Equation \eqref{eq: leapfrog} or \eqref{eq:MN}.\;
}
\end{algorithm}

One option is to introduce occasional billiard-like bounces \citep{billiards,BIoptimization}. At the bounce the momentum is suddenly reoriented to a new, isotopically randomly chosen direction, while 
conserving the momentum length and 
thus the energy. The MCHMC pseudocode algorithm for a variable mass Hamiltonian with bounces is shown in Algorithm \ref{alg: bounces}.
The frequency of the bounces is a hyperparameter, which can significantly influence the sampling efficiency, see Figure \ref{fig: bounces}. It is analogous to the frequency of the momentum resampling in HMC, and can be tuned by a preliminary run \citep{NealHandbook}. In Section \ref{sec: tuning} we will present a tuning-free scheme.


\begin{figure}
    \makebox[\textwidth][c]{\includegraphics[scale = 0.32]{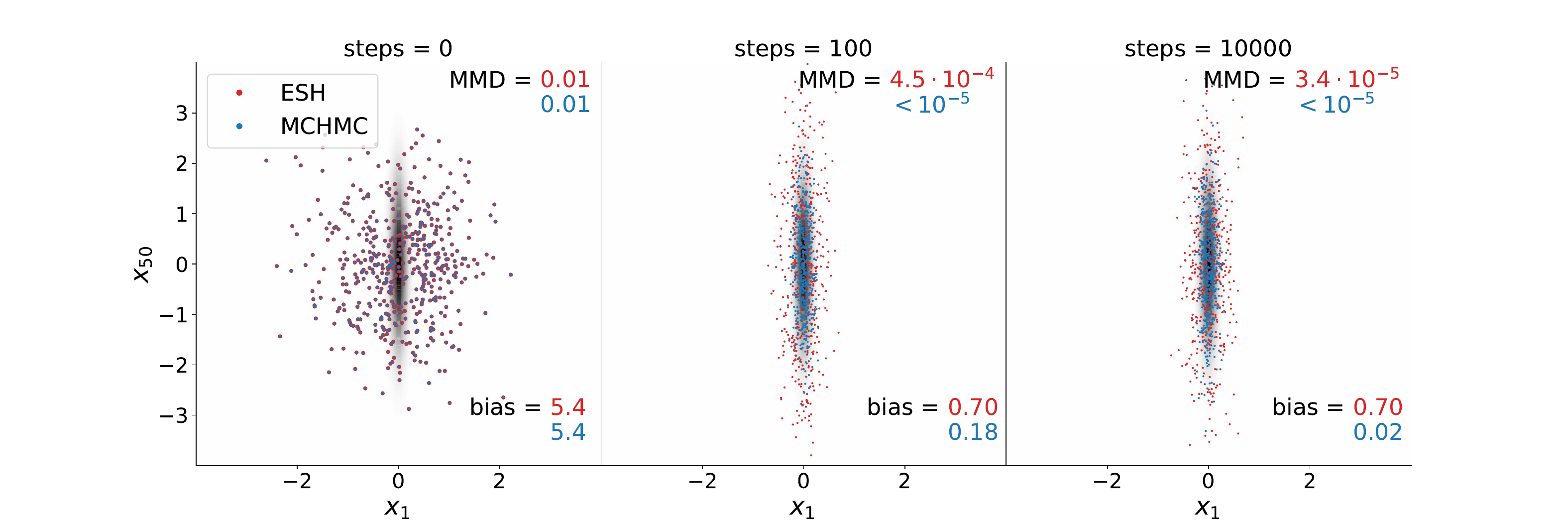}}
    \caption{We sample from the 50D ICG example \citep{ESH}: a 50-dimensional Gaussian with the condition number $\kappa = 100$ and variances arranged linearly between 0.01 and 1. The same setup as in \cite{ESH} is used: we run 500 parallel chains, each initialized from a standard Gaussian. The integration stepsize is $\epsilon = 0.5$. Only the largest and shortest eigenvalue directions are shown. Particle snapshot after 0 (initial condition), 100 and 10000 steps is shown. The target distribution density is shown in gray. The ESH dynamics without bounces (red) never converges, unlike MCHMC (blue). In \cite{ESH}, maximum mean discrepancy (MMD) was used to measure posterior quality. We show MMD at 0, 100 and 10000 steps for ESH (red) and MCHMC (blue). MMD requires very low values for convergence, and MMD calculation implementation \citep{ESH} fails if MMD is very low and returns negative values, in which case we state $MMD < 10^{-5}$. ESH convergence failure is easily revealed with bias, which settles at $b_2 = 0.70$. On the contrary, MCHMC is very close to the true posterior after 10000 steps, with $b_2 = 0.02$.} 
    \label{fig: ESHexample}
\end{figure}

The importance of bounces is exhibited
in Figure \ref{fig: ESHexample}. Here 
we run 500 chains in parallel. We 
compare no bounce ESH algorithm \citep{ESH} to MCHMC with tuning-free bounce
frequency on a 50-dimensional 
ill-conditioned Gaussian target. 
We initialize from a broad prior (steps=0). We observe that without bounces, the ESH algorithm never relaxes
the particles into the target distribution. 
In contrast, MCHMC particles are closer to the target even at 100 steps, 
and even more so at 10000 steps. We do the bounces with the tuning-free prescription
described below.
We also quote Maximum Mean Discrepancy (MMD) used in \cite{ESH}, which requires very low values
for convergence on posterior.
As such, it does not directly relate to the quality of 1d marginal posteriors, which is 
our metric of quality.

\subsection{Microcanonical Langevin-like Monte Carlo} \label{sec: MLMC}

\begin{algorithm}
\caption{MCLMC $q=0$ algorithm.}\label{alg: Langevin}
\KwData{
$\text{initial condition } \x_0 \in \mathbb{R}^d$, \\
$\text{number of samples } N \in \mathbb{N}$,
$\text{step size } \epsilon > 0$, \\
$\text{decay constant } L > 0$.
}
\KwResult{$\text{samples} \{ \x_n \}_{n = 1}^N$, $\text{weights} \{ w_n \}_{n = 1}^N$}
$w_0 \gets 1$\;

\For{$n\gets0$ \KwTo $N$}{  
    $\x_{n+1}, \, \U_{n+1}, \, w_{n+1} \gets 
 \big(\Phi_{\epsilon, L}^{O}\circ \Phi_{\epsilon} \big)(\x_n, \, \U_n, \, w_n)$, see Equation \eqref{eq: MCLMC update}.;
 
}
\end{algorithm}

Langevin Monte Carlo follows the underdamped Langevin dynamics, which  for a 
standard kinetic energy can be written as 
\begin{equation}
    \dot{\x} = \p, \qquad \dot{\p} = - \nabla V(\x) -\gamma \p +(2\gamma)^{1/2}\frac{d{\bf W}}{dt}, 
\end{equation}
where ${\bf W}$ is the white Gaussian noise
and $\gamma>0$ is friction coefficient
\citep{MolecularDynamics}, Many standard discretization schemes 
are a combination of deterministic coordinate ($\Phi^T$) and momentum ($\Phi^V$) updates, together with a stochastic partial momentum refreshment update:
\begin{equation}
    \Phi^O_{\epsilon, \gamma}(\x, \p) = (\x, \, \eta \p+(1-\eta^2)^{1/2}\boldsymbol{z}),
    \label{Gn}
\end{equation}
where $z_i \sim \mathcal{N}(0, 1)$
and $\eta=\exp(-\gamma \epsilon)$, 
where $\epsilon$ is the stepsize.  
For example, velocity leapfrog update corresponds to 
$\Phi^{O}_{\epsilon} \circ \Phi^{V}_{\epsilon/2} \circ \Phi^T_{\epsilon} \circ \Phi^{V}_{\epsilon/2}$ scheme.

In contrast to Langevin dynamics, MCHMC 
is energy conserving, and for $q=0$
has a non-standard kinetic term. 
We develop the analogous
expressions in the rescaled time 
formulation:
\begin{equation} \label{eq: MCLMC update}
    (\x_{n+1},\, \U_{n+1}, w_{n+1}) =  ( \Phi_{\epsilon, L}^{O}\circ \Phi_{\epsilon})(\x_n, \, \U_n, \, w_n).
\end{equation}
Here, $\Phi_{\epsilon}$ simulates Equation \eqref{eq: rescaled EOM} by either leapfrog \eqref{eq: leapfrog} or minimal norm integrator \eqref{eq:MN}.
The $\Phi^O_{\epsilon, L}$ is a partial momentum direction refreshment, which preserves momentum magnitude and therefore the energy. We adopt
\begin{equation} \label{eq: partial refreshment}
    \Phi^O_{\epsilon, L}(\x, \U, w) = (\x, \frac{\boldsymbol{u} + \nu \boldsymbol{z}}{ \vert \boldsymbol{u} + \nu \boldsymbol{z} \vert}, w).
\end{equation}
with
\begin{equation}
    \nu = \sqrt{\frac{1}{d} \big( e^{2 \epsilon / L} - 1\big)}
\end{equation}

Note that $\nu$ plays a similar role as $(1 - \eta^2)^{1/2}/\eta$ in Equation \eqref{Gn}. 
We observe that the noise we add
is non-Gaussian and that the coefficient in front of the 
drift term $\boldsymbol{u}$ is 
stochastic as it depends on $\boldsymbol{z}$. Non-Gaussian Langevin-like 
dynamics has been investigated in physics 
literature \citep{PhysRevLett.114.090601}. 

As we will show below, $L$ can be interpreted as a distance after which the repeated application of \eqref{eq: partial refreshment} builds up to give a complete decoherence of the momentum, an effect similar to a random bounce. Therefore, both the MCHMC with bounces and MCLMC share parameters with a similar meaning. 

For small $\nu$ in high dimensions, one application of \eqref{eq: partial refreshment} rotates the momentum by an angle $\cos \alpha = (1 + \tan^2 \alpha)^{-1/2} \approx (1 + \nu^2 d)^{-1/2}$. This is because the change of $\boldsymbol{u}$ in the direction perpendicular to $\boldsymbol{u}$ is Gaussian distributed and in high dimensions lies on its typical set (not to be confused with the typical set of the target distribution in the configuration space). Therefore, its magnitude is $\nu \sqrt{d -1} \approx \nu \sqrt{d}$. 

The momentum correlations then decay exponentially with the number of steps $n$:
\begin{equation} \label{eq: correlations}
    \langle \boldsymbol{u}_n \cdot \boldsymbol{u}_0 \rangle = (1 + \nu^2 d)^{-n / 2} = e^{- n \epsilon / L},
\end{equation}
and $L$ is the decay distance. For a proof, see \citet{SphereRandomWalk}.

In the limit of a large $\nu$, MCLMC does 
full momentum refreshments, and is equivalent to MCHMC 
with $L = \epsilon$, analogous to overdamped 
Langevin dynamics, which is known 
to correspond to a single leapfrog step 
of HMC \citep{riemannHMC}. This choice 
is however suboptimal (Figure \ref{fig: bounces}).
A hybrid between HMC and LMC is 
generalized HMC, which adds some small random element to the momentum after $K$ 
Hamiltonian dynamics steps, rather than completely resampling the momentum occasionally as in HMC, or rather than 
partially resampling it at every step as in LMC \citep{generalizedHMC, meads}. \citet{generalizedHMCproof} shows the superiority of this strategy on the ill-conditioned targets. We do not 
investigate this further here in the 
context of MCHMC and MCLMC.

\begin{figure}
    \makebox[\textwidth][c]{\includegraphics[scale = 0.17]{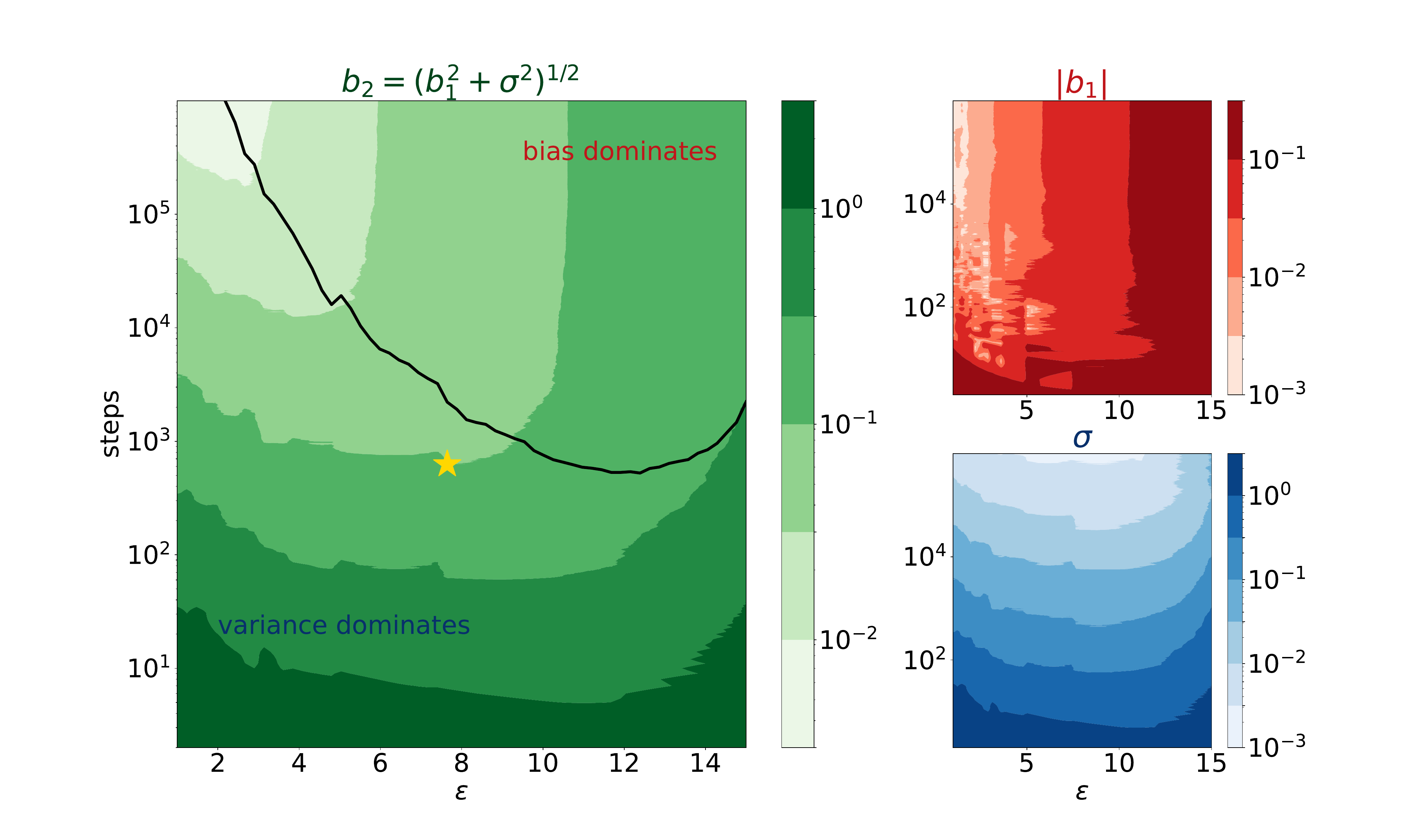}}
    \caption{Bias and variance for $d=100$ standard Gaussian. We show the absolute bias (upper right) and variance (lower right) as a function of step size $\epsilon$ and the number of steps. The combination $b_2=({b_1^2+\sigma^2)^{1/2}}$ is shown on the left. The curve where the contributions from the bias and the variance are equal is shown in black. The yellow star denotes the optimal setting of $\epsilon$ if one is aiming for $b_2 = 0.1$.}
    \label{fig:b}
\end{figure}

\subsection{Bias versus variance}

In the Bayesian analysis, a sampler is typically used to estimate the uncertainty region of the marginal posterior of parameters. Quantities of direct interest are therefore the relative errors of the second moments:
\begin{equation}
    z_i = \frac{\mathbb{E}_{\text{sampler}} [x_i^2] - \mathbb{E}_{\text{truth}} [x_i^2]}{\mathbb{E}_{\text{truth}} [x_i^2]}.
\end{equation}
We define the dimension-averaged error on the first moment, or simply bias ($b_1$), dimension-averaged error on the second moment ($b_2$) and variance over the dimensions ($\sigma^2$):
\begin{equation}
    b_1 = \langle z_{i} \rangle \qquad \sigma^2 = \langle (z_{i} - b_1 )^2 \rangle \qquad b_2^2 = \langle z_i^2 \rangle = b_1^2 + \sigma^2,
    \label{eq:b}
\end{equation}
where $\langle \cdot \rangle = \frac{1}{d} \sum_{i = 1}^d$ is the average over the dimensions. We judge the quality of the sampler based on how many target density gradient evaluations it needs to get the average second moment error $b_2$ below a predefined threshold \citep{UsingBias, DLMC}. MCHMC without bounces is an Ordinary Differential Equation (ODE) solver of Hamiltonian dynamics, and its accuracy is controlled by the step size $\epsilon$. The length of trajectory and frequency of bounces  give rise to a stochastic component to the error, which is decreased as the trajectory length increases. Thus, MCHMC will in general have both bias and variance. In HMC literature, it is common to insist on unbiased estimators, which can be accomplished by performing a Metropolis acceptance or rejection after the integration (e.g. \cite{conceptualHMC}). However, as long as the $b_1^2$ can be controlled to be below the 
variance $\sigma^2$ this step is unnecessary, and indeed it is not used in fields such as Molecular Dynamics \citep{MolecularDynamics}.

In Figure \ref{fig:b} we explore the dependence of bias and variance on step size $\epsilon$ and number of steps  for a $d=100$ standard Gaussian. The bias (upper right) depends only on step size $\epsilon$. 
The bottom right plot shows the scaling of square root variance with $\epsilon$ and number of steps $N_{\rm steps}$. For most of the regime below some critical value of $\epsilon$ the variance scales roughly as ${\rm variance} \propto N_{\rm steps}^{-1}\epsilon $. The scaling of variance inversely with $N_{\rm steps}$ is expected, since the number of effective samples scales linearly with $N_{\rm steps}$. We also expect that we obtain fewer effective samples if we reduce $\epsilon$ below its critical value, since the distance travelled over a fixed number of steps is shorter, although the number of steps between two bounces $L$ also plays a role in determining ESS. Above the critical stepsize $\epsilon \sim 12$ the integrator becomes unstable, and we accumulate both large bias and variance. 

The left panel combines the two errors. We also show the 
 $b_1^2=\sigma^2$ line where the two contributions to 
$b_2^2$ are equal, which is expected to 
be close to the optimal choice (star). We see that 
the optimal choice of the stepsize and number
of steps depends on the requirements for $b_2$: we need more than 300 times more steps if we 
want to reach $b_2=0.01$ than for $b_2=0.1$. However, the requirements for $\epsilon$ are less 
stringent: for $b_2=0.1$ the optimal choice is 
$\epsilon=8$, and for $b_2=0.01$ it is closer 
to $\epsilon=3$. We observe that $b_1^2$, which 
enters Equation \eqref{eq:b}, scales as $\epsilon^4$. 
Thus, reducing $\epsilon$ by a factor of 2 
we reduce the contribution of $b_1$ term to $b_2$
by a factor of 16, while only paying about 
a factor of 2 in computational cost due to the larger required 
number of steps. Thus, even a small change of 
$\epsilon$ can make the bias contribution to the 
overall error completely negligible. This is also 
the reason for the optimal value of $\epsilon$ being lower
than the $b_1^2=\sigma^2$ line.

\subsection{Hyperparameter tuning} \label{sec: tuning}

We here design an efficient algorithm for tuning the  integration stepsize $\epsilon$ and the momentum correlation decay length $L$.

\subsubsection{Integration stepsize $\epsilon$}

\begin{figure}
    \makebox[\textwidth][c]{\includegraphics[scale = 0.29]{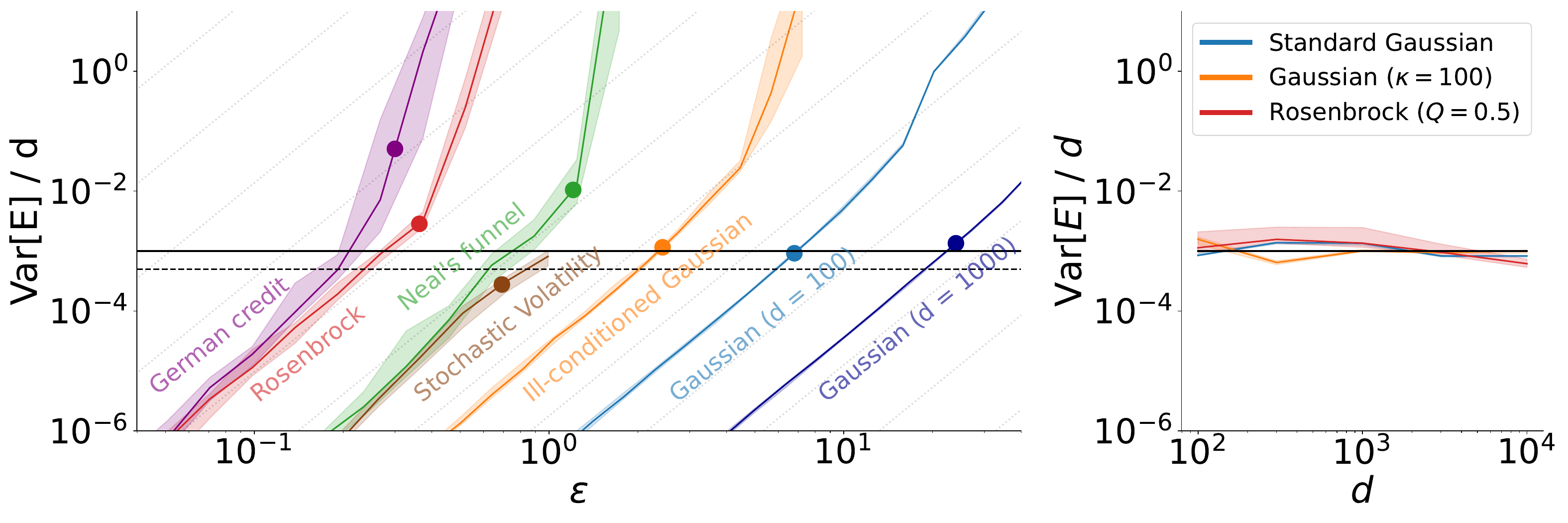}}
    \caption{Left: Energy variance per dimension Var[$E$]/$d$ is shown against step size $\epsilon$. The solid lines are the medians over the random seeds, the confidence bands are the first quartiles. We show various target distributions from Section \ref{sec: results}. We use the MLMC algorithm with the leapfrog integrator and the optimal $L$. For $\mathrm{Var}[E]/d <0.001$ the typical scaling is $\mathrm{Var}[E]/d \propto \epsilon^4$. The $\epsilon^4$ lines are shown in the background. The optimal $\epsilon$ from the grid search is shown by a circle.
    The choice of $\mathrm{Var}[E]/ d\sim 0.001$
    is close to optimal for all cases (solid black line). The conservative choice $0.0005$ of the tuning algorithm is shown with a dotted line. Right: scaling of Var[$E$]/$d$ with $d$ at optimal $\epsilon$: it is constantly around $0.001$. }
    \label{fig:energy}
\end{figure}

Since we do 
not have Metropolis adjustment, we 
must control the bias with the 
step size. 
Symplectic leapfrog integrators
and its relatives do not 
accumulate the energy error over
many orbits, and the error 
remains constant and of order 
$\epsilon^2$ as long as $\epsilon$
is below the critical value, above 
which instability occurs such 
that the kinetic energy is 
no longer positive definite. 
Here we argue that in MCHMC, which 
is energy conserving, 
bias can be estimated
by monitoring energy fluctuations. 
We define $\mathrm{Var}[E]$ as the mean 
square energy fluctuations during the 
integration of last equation \eqref{eq: rescaled EOM}.

At the optimal hyperparameters, we would expect $\text{Var}[E] \propto d$ if sampling along each coordinate was independent. In MCHMC the dynamics are coupled, but the right panel of Figure \ref{fig:energy} shows that $\text{Var}[E] \propto d$ is still true. The Figure \ref{fig:energy} also shows that the optimal stepsize setting corresponds to $\mathrm{Var}[E]/d \sim 0.001$. This is true for various benchmark target distributions (defined in Section \ref{sec: results}). We can thus tune $\epsilon$ to achieve desired energy fluctuations per dimension, and this in turn guarantees low bias on the posteriors. 
Note that the dependence of the energy fluctuation on the stepsize is very steep ($\text{Var}[E] \propto \epsilon^4$) so the stepsize is not very sensitive to the choice $0.001$. We will target $\mathrm{Var}[E]/d \sim 0.0003$ as a conservative choice. We do a short run with a few hundred steps and $\epsilon_0 = 0.5$ to determine $\mathrm{Var}[E]$ and update the stepsize to $\epsilon = \epsilon_0 (0.0005 \, d / \mathrm{Var}[E])^{1/4}$. We repeat this step a few times for convergence.

One could also monitor the fluctuations of the weights $w$. We observe that the weights are almost nearly 
constant up to the critical $\epsilon$ (of order 12 for the standard Gaussian example of Figure \ref{fig:energy}), beyond which their fluctuations significantly increase. Thus, the weights can inform us of the onset of integration instability, although similar information is also obtained from $\mathrm{Var}[E]/d$.

\subsubsection{Momementum decoherence scale $L$}

\begin{figure}
    \makebox[\textwidth][c]{\includegraphics[scale = 0.3]{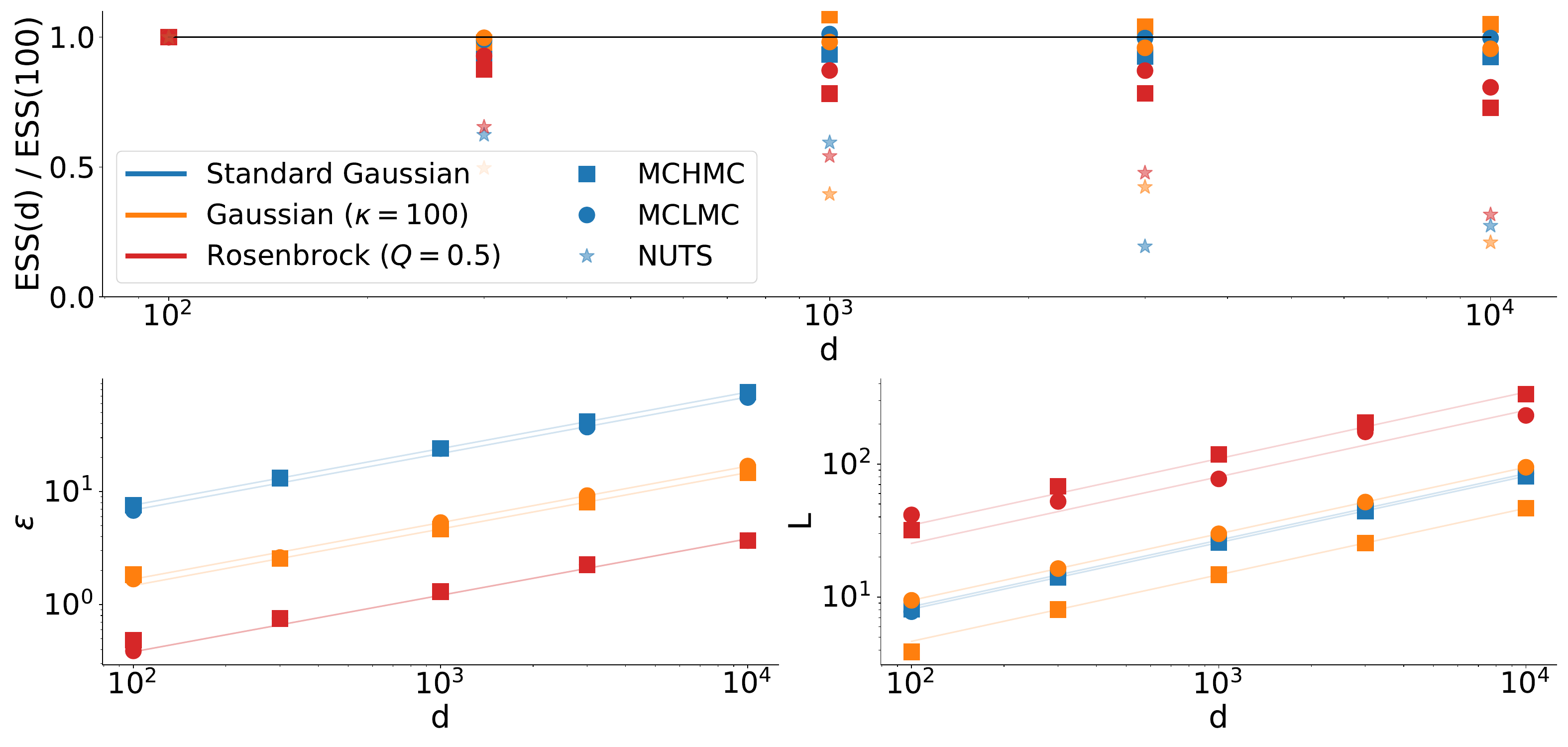}}
    \caption{We study the optimal hyperparameter $\epsilon$ and $L$ settings as a function of the target dimension $d$. We show three types of targets: Standard Gaussian (blue), Ill-conditioned Gaussian with $\kappa = 100$ (orange) and the Rosenbrock function with $Q = 0.5$ (red). All targets are normalized to the unit dimension-averaged variance $\sigma^2_{\text{eff}} = 1$. We study both standard MCHMC (squares) and MCLMC (circles). We use $q = 0$ Hamiltonian and the leapfrog integrator.
    For each target, we do a grid search over the hyperparameters $\epsilon$ and $L$ to determine the values which maximize the ESS. The optimal $\epsilon$ as a function of the target dimension is shown in the bottom left panel. The optimal $L$ is shown in the bottom right panel. The ESS relative to the ESS in $d = 100$ is shown in the upper panel. Importantly, the optimal ESS is independent of the dimension. In all cases, both $\epsilon$ and $L$ scale as $\sqrt{d}$, as expected. The best fit lines are shown. $\alpha = L / (\sigma_{\text{eff}} \sqrt{d})$ is on the order of unity.
    }
    \label{fig: scaling}
\end{figure}

The typical set of an arbitrary distribution $p(\x)$ is as a set of all $\x$ whose $-\log p(\x)$ is close to the entropy of the given distribution $S_p = - \int p(\x) \log p(\x) d \x $. The probability of the sampler being in a small neighborhood of the typical set approaches unity for large $d$ \citep{typical_set1, typical_set2}.
We expect the optimal decay length of the momentum correlations $L$ to be on the same scale as the typical set, because we do not want the sampler to be caught in orbits, similar to the No-U-Turn condition in NUTS \citep{NUTS}.

The typical set of the Gaussian with an isotropic covariance matrix $\sigma^2 I$ is a sphere of radius $\sigma d^{1/2}$. We therefore expect the optimal $L$ for a Gaussian to be $L = \alpha \sigma d^{1/2}$, where $\alpha$ is a constant of order unity. Figure \ref{fig: scaling} confirms this intuition and shows $\alpha \approx 1$.

If the target is non-Gaussian the simplest approximation is to generalize $L = \sigma_{\text{eff}} \sqrt{d}$ and estimate the effective width of the posterior as a dimension averaged variance: $\sigma_{\text{eff}}^2 = \frac{1}{d} \sum_{i = 1}^d \text{Var}[x_i^2]$. We determine $\sigma_{\text{eff}}$ as a side product in the stepsize tuning.

However, for the highly non-Gaussian targets, the geometry of the typical set differs from the sphere and the above approach becomes suboptimal. An example is the Rosenbrock function target, where the required $\alpha \approx 4 \neq 1$ (see Figure \ref{fig: scaling}). 
As a more general approach, we use $L = \sigma_{\text{eff}} \sqrt{d}$ as an initial approximation and run the sampler for $n$ steps to determine the effective sample size $n_{\text{eff}}^{(i)}$ for each parameter $x_i$ using the autocorrelations \citep{blackjax, autocorr, autocorr2}. The distance between the effective samples is then 
\begin{equation}
    l = \frac{\epsilon}{d^{-1} \sum_{i = 1}^d n_{\text{eff}}^{(i)} / n}.
\end{equation}
The optimal momentum decoherence length should be on the same scale, we find that $L = 0.4 \, l$ works well for all benchmark problems in this paper.

The required number of steps $n$ in the preliminary run scales with the difficulty of the problem, because the distance between the effective samples is larger, and we need more steps to compute it. However, $n$ is always considerably lower than the number of steps needed for convergence to $b_2 = 0.1$. On the benchmark problems tested in this paper, $n > 10 l / \epsilon$ was a sufficient criterion for convergence. This gives for example $n \approx 140$ for the Ill-conditioned Gaussian example and $n \approx 600$ for the Stochastic Volatility example.  

More sophisticated tuning algorithms using jumping distance optimization \citep{NUTS} or Change in the Estimator of the Expected Square (ChEES) \citep{ChEES} are likely to improve further this simple tuning scheme, at the expense of higher computational cost, which might not be justified, given the close to optimal performance of the presented algorithm (compare Tables \ref{table} and \ref{table2}).
When possible, another promising path is to tune optimal 
parameters in low dimensions and use their scaling $\epsilon \propto d^{1/2}$ and $L \propto d^{1/2}$ 
of Figure \ref{fig: scaling} to extend to higher dimensions.

\subsection{Related work} 
While the samplers we propose are based on 
Hamiltonian dynamics, their underlying 
justification differs from the 
standard HMC \citep{HMCDuane,NealHandbook}. HMC relies on detailed balance arguments, which require occasional stochastic momentum resamplings that change
the energy of the system. Instead, MCHMC 
is energy conserving for the entire trajectory, and we need to additionally
assume we explore all the microcanonical 
states on that energy surface. 
A specific model (ESH) from the MCHMC class of models was recently proposed in \citet{ESH}, where an efficient numerical integrator for the proposed Hamiltonian was also developed.  This work assumes that the ergodic hypothesis holds when averaged over many independent chains, such that the microcanonical ensemble average equals the time average over the Hamiltonian trajectories, which we show 
does not hold in practice. 


While stochastic momentum 
resamplings are essential for validity of HMC, 
we show they are also essential for 
MCHMC, in that 
ergodicity is not achieved without 
the bounces, regardless of 
whether one uses one or multiple 
chains. Billiard-like bounces,
which randomize momentum while conserving energy,
achieve 
chaotic mixing of the orbits,  and were  introduced in the optimization context in \citet{BIoptimization} to encourage rapid phase space exploration. 
We compare full occasional momentum refreshments against
partial momentum refreshments at every step,
which are used in underdamped Langevin 
dynamics \citep{MolecularDynamics}. 
MCHMC and MCLMC do not have Metropolis
adjustment, and are thus related
to unadjusted Langevin and 
Hamiltonian MC, which are 
common in fields such as 
Molecular Dynamics \citep{MolecularDynamics}. All the
unadjusted
methods have a bias, which 
must be controlled to be lower
than the variance.

\section{Experiments} \label{sec: results}

\begin{table}[]
    \makebox[\textwidth][c]{\begin{tabular}{ccccccc}
    \toprule
           & Ill-conditioned & Bi-modal& Rosenbrock & Neal's &  German & Stochastic \\ 
           & Gaussian & &  & Funnel &  Credit & Volatility    \\\midrule
    Langevin-like, MN, $q = 0$ & \textbf{0.110} & \textbf{0.064} & 0.0033 & \textbf{0.021} & 0.0099 & \textbf{0.023}\\
    Langevin-like, LF, $q = 0$ & 0.075 & 0.047 & 0.0031 & 0.013 & \textbf{0.0105} & 0.016\\ 
    bounces, LF, $q = 0$ & 0.039 & 0.039& 0.0030 & 0.021 & 0.0040 & 0.014\\
    \midrule
    bounces, Y, $q = 2$&  0.025  & 0.041 & 0.0005 &  \, 0.0001 &  0.0003 & 0.00001\\ \midrule
    no bounce, LF, $q = 0$ & 0 & 0 & 0.0012 & 0 & 0.0019 & 0\\
    ESH, LF, $q = 0$ & 0 & 0 & 0.0004 & 0.001 & 0.0003 & 0 \\ \midrule
    NUTS   & 0.012 & 0.008 & 0.0015 & 0.006  & 0.0014& 0.006\\
    unadjusted HMC & 0.031 & 0.019 & \textbf{0.0051} & 0.004 & 0.0025 & 0.002 \\
    \end{tabular}}
    
    \caption{Sampling efficiency (ESS, see \eqref{ESS definition}) comparison between various tuned versions of MCHMC, unadjusted HMC and NUTS, where the tuning is not included in the sampling cost. Higher is better, the best performers are shown in bold.  
    The first column indicates the momentum decoherence mechanism (bounces, Langevin-like, or no bounce), the integrator used (leapfrog (LF), minimal norm (MN) or Yoshida (Y) and the Hamiltonian used ($q = 0$ or $q = 2$).
    $q= 0$ Hamiltonian gives the best results and significantly outperforms NUTS. Different decoherence mechanisms achieve comparable efficiency on most targets.
    Minimal norm integrator typically outperforms the leapfrog integrator.
    Algorithms without momentum decoherence are suboptimal, and fail to converge to the desired accuracy on several examples, regardless of whether we run a single chain (no bounce), 
    or multiple chains with different initial conditions (ESH).
    }
    \label{table}
\end{table}

\begin{table}[]
    \makebox[\textwidth][c]{\begin{tabular}{cccccccc}
    \toprule
           & Ill-conditioned & Bi-modal& Rosenbrock & Neal's &  German & Stochastic & Cauchy\\ 
           & Gaussian & &  & Funnel &  Credit & Volatility  &  \\\midrule
    MCLMC & \textbf{0.075} & \textbf{0.045}& \textbf{0.0021} & \textbf{0.0078} & \textbf{0.0059} & \textbf{0.011} & \textbf{0.001}\\ \midrule
    NUTS  & 0.006 & 0.006 & 0.0008 & 0.0019 & 0.0008& 0.001 & $< 10^{-6}$\\ 
    \end{tabular}}
    \caption{Sampling efficiency (ESS) comparison between automatic tuning MCLMC and NUTS (where the tuning run of 500 steps is included in the sampling cost). Here we use the leapfrog integrator with $q = 0$. Higher is better, the best performers are shown in bold. Note that different ESS definition was used for the Cauchy example because of the diverging second moments, see Section \ref{sec: cauchy}.
    }
    \label{table2}
\end{table}

We use the Gaussian distribution with 
zero mean to define $b_2^2 \equiv 2 / n_{\text{eff}}$, where $n_{\text{eff}}$ is the effective number of independent samples. Typically, this is the lowest possible 
number of samples needed to achieve the 
target value of $b_2$, and for targets with fat tails it takes significantly larger $n_{\text{eff}}$
to reach a given value of $b_2$.
We define the effective sample size (ESS) as an effective number of samples produced per target gradient evaluation \citep{UsingBias, DLMC}.
For correlated MCMC chains with equal weight, 
ESS is often defined in terms of 
correlation length, while for uncorrelated samples with importance weights ESS is 
defined in terms of weight fluctuations. 
In our case we have both weights and 
correlations, so we choose to define 
ESS on the quantity that is relevant 
for the quality of posteriors, which 
is the error of the second moment. 
We take the threshold $b_2 = 0.1$, 
which typically corresponds to a 
notion of a converged posterior. 
Adopting this case, we define
\begin{equation} \label{ESS definition}
    \text{ESS} \equiv \frac{200}{n},
\end{equation}
where $n$ is the number of target density gradient evaluations at which $b_2 = 0.1$. 
The reported ESS in this paper are averages over $10$ random seeds and initial condition draws. The initial condition is drawn at random from the prior, which is a standard Gaussian (except in the Stochastic Volatility example). We have verified that our ESS
definition broadly agrees with the 
standard definition of ESS via the 
correlation length of the chain.

We will envision two common scenarios for the comparison against the baselines. In the first scenario, we first perform a grid search over the hyperparameters, and then evaluate ESS without counting the tuning in the cost. This corresponds to the practical situation where the sampler is tuned only once and then used repeatedly on similar problems. We present the results in Table \ref{table}. 
In the second scenario, the cost of tuning is included in the ESS. The results are shown in \ref{table2}. Here, we use an efficient tuning algorithm from Section \ref{sec: tuning} (Table \ref{table2}). 
Our default MCHMC choice is $q=0$, so whenever not specified we refer to $q=0$, but we also ran $q= 2$ case. 

We compare MCHMC against the state-of-the-art variant of the HMC, the No-U-Turn Sampler (NUTS) \citep{NUTS} as implemented in the NumPyro library \citep{NummPyro}. This sampler requires a warm-up pre-run to adjust the integration step size and the mass matrix. We use the recommended warm-up of 500 HMC samples, and we present results without warm-up in the ESS 
 in Table \ref{table} and Figure \ref{fig:kappa}, and with 
 warm-up in Table \ref{table2}. 
Since MCHMC is an unadjusted 
method, we also compare it 
to unadjusted HMC for the case 
where we optimize the hyperparameters. In this case, 
unadjusted HMC sometimes 
outperforms NUTS (Table \ref{table}), 
but its results are very sensitive to 
the tuning. 
We were unable to
find a good tuning-free solution 
for unadjusted HMC, and we do 
not show it in Table \ref{table2}. 
We now define the benchmark problems.


\subsection{Ill-conditioned Gaussian}
        
    \begin{figure}
        \makebox[\textwidth][c]{\includegraphics[scale = 0.25]{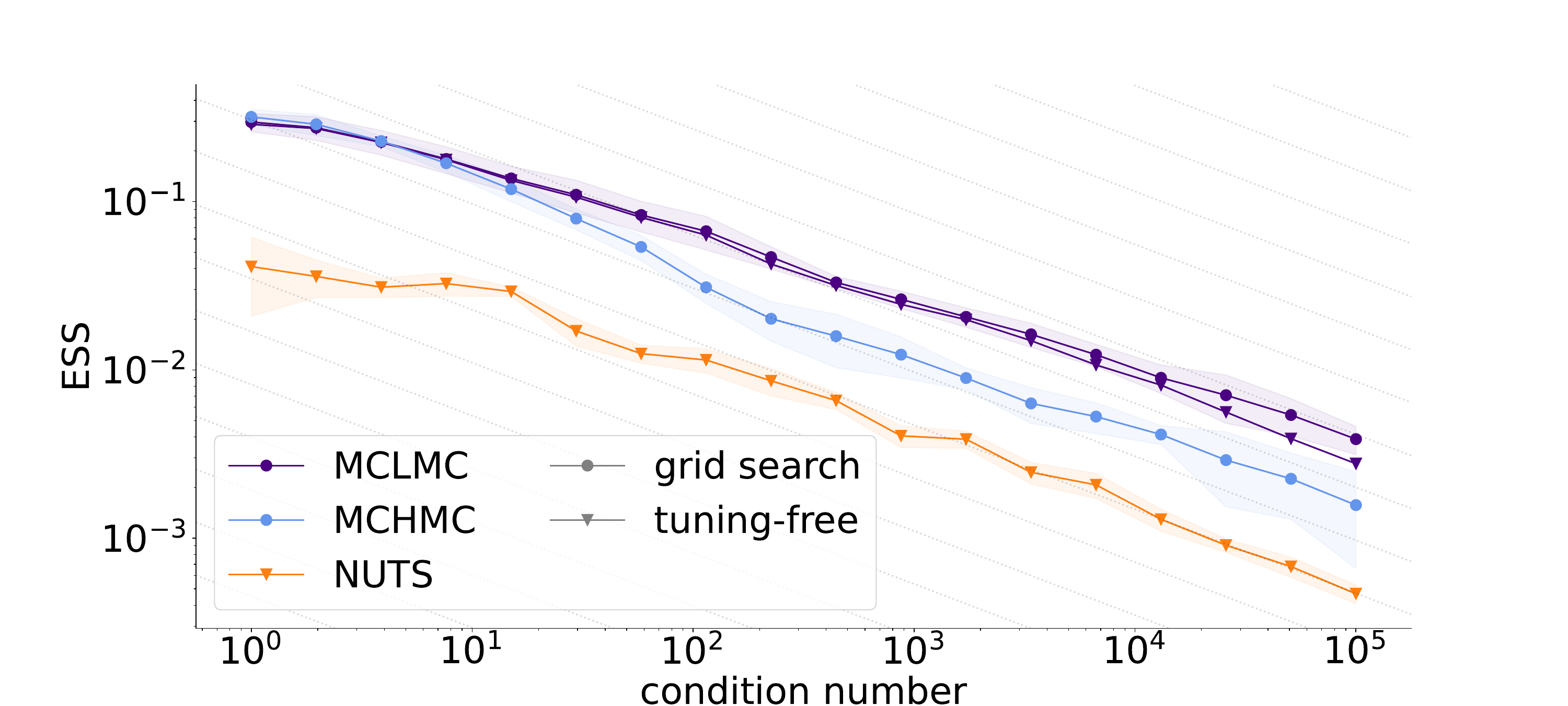}}
        \caption{We compare the ESS performance of MCHMC $q=0$ with leapfrog integrator to NUTS on $d=100$ ill-conditioned Gaussians. 
        The reported confidence bands are the standard deviations over the prior realizations, not the uncertainties of the average (which are by a factor of 3 smaller). They indicate how much we should expect our results to vary from run to run. 
        Using our tuning-free algorithm (triangles)
        is practically optimal.
        NUTS results do not count warm-up, which can increase its computational cost significantly. The ${\rm ESS} \propto \kappa^{-1/2}$ lines are shown in gray, and we see that MCLMC achieves a shallower dependence ${\rm ESS} \propto \kappa^{-0.38}$.}
        \label{fig:kappa}
    \end{figure}
    
    This  is a 100-dimensional Gaussian with a high condition number $\kappa$ of the covariance matrix. We take a randomly orientated covariance matrix with eigenvalues equally spaced in log between $1/\sqrt{\kappa}$ and $\sqrt{\kappa}$. 
    We compute the ESS in the coordinates in which the covariance matrix is diagonal and take the analytical ground truth second moments. The results are shown in Figure \ref{fig:kappa}. In Table \ref{table} we report the results for $\kappa = 100$. We see that at $\kappa=100$ MCLMC outperforms NUTS (with warm-up) by more than an order of magnitude, and this improvement 
    further increases for higher 
    condition numbers. Even at 
    condition number of 1 the improvement  is a factor of 5. 
    MCHMC without bounces and ESH fail to converge
    on this example because it is 
    not ergodic due to the symmetries
    of the Gaussian distribution.

\subsection{Bi-modal distribution} 
        
    \begin{figure}
        \centering
        \begin{minipage}{0.45\textwidth}
            \hspace*{-1cm}\includegraphics[scale=0.24]{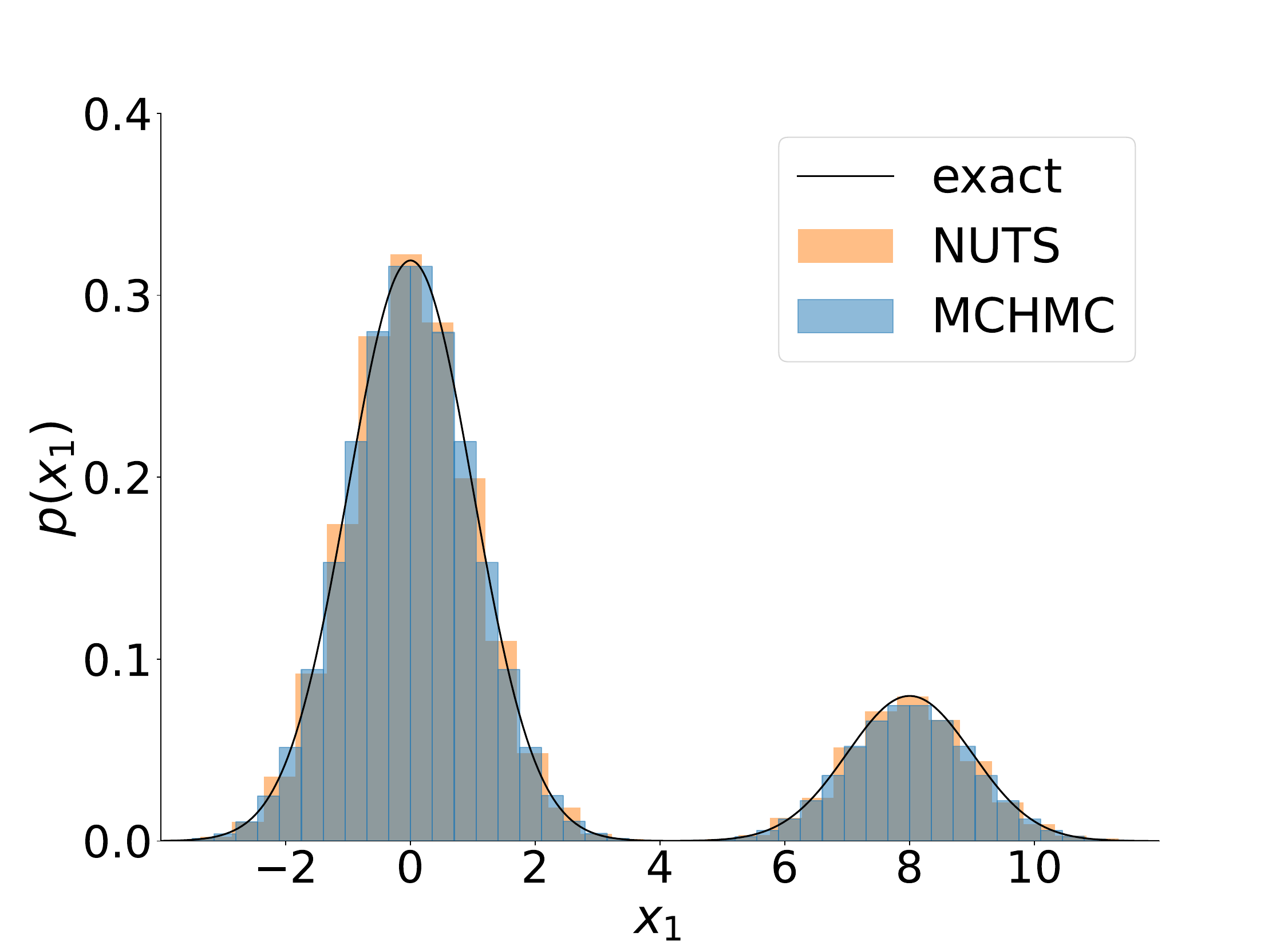}
        \end{minipage}
        \begin{minipage}{0.45\textwidth}
            \hspace*{1.7cm}\includegraphics[scale=0.23]{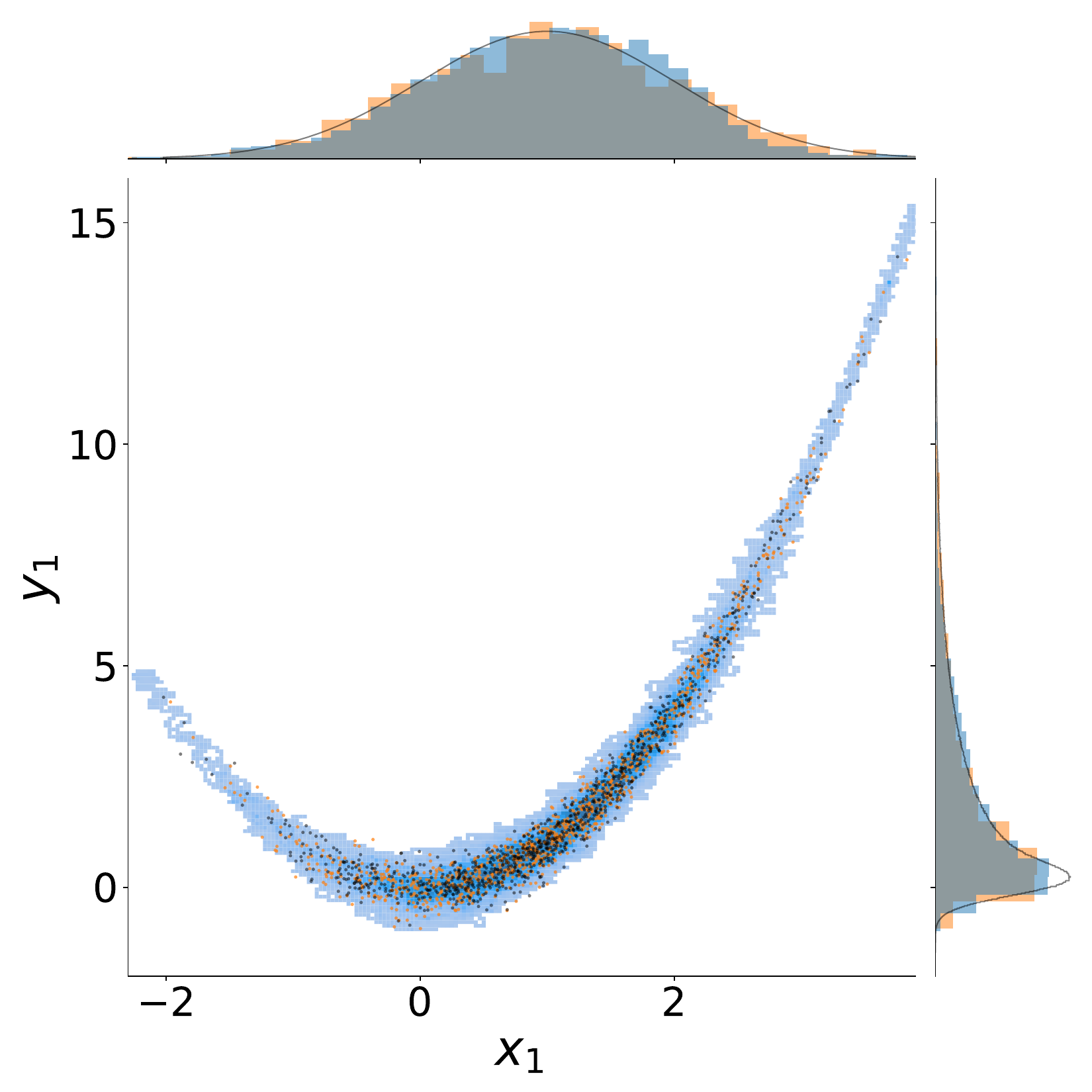}
        \end{minipage}
        \caption{Left: the target is a 50-dimensional 80 \% - 20 \% Gaussian mixture with two modes, separated by $8 \sigma$. We show the marginal distribution along the mode separation axis. Both NUTS and MCHMC give an accurate posterior after a very long run ($10^7$ steps).
        Right: 32-dimensional Rosenbrock target with $Q = 0.1$. We show the marginal distribution in the ($x_1$, $y_1$) plane and the one-dimensional marginals, computed with MCHMC (blue), NUTS (orange), and by the generating process (black). Both samplers accurately capture the target.}
        \label{fig:rosenbrock}
    \end{figure}
    
    This is an 80 \% - 20 \% mixture of two standard 50-dimensional Gaussians, separated by $8 \sigma$. Note that their typical sets are still close, since $\sqrt{50}\simeq 7$. The ground truth moments are known analytically. The posterior along the separation axis is shown in Figure \ref{fig:rosenbrock}. Table \ref{table} shows that MCHMC improvement over NUTS (with warm-up) is a factor of 6-10.
    MCHMC without bounces and ESH fail to converge
    on this example.

\subsection{Rosenbrock function}
    
    This target    has a narrow banana shape, designed to be a problematic test example \citep{rosenbrock}. We take $d/2 = 18$ independent copies of two-dimensional bananas in $(x_i, y_i)$ spaces \citep{DLMC}:
    
    \begin{equation} \nonumber
        p(\x, \boldsymbol{y}) = \prod_{i = 1}^{d/2} \mathcal{N}(x_i \vert 1, 1) \, \mathcal{N}(y_i \vert x_i^2, Q^{1/2}).
    \end{equation}
    
    Here, $\mathcal{N}(x \vert \mu, \sigma)$ is the Gaussian probability density distribution and $Q = 0.1$ is a parameter determining the width of the bananas. $\langle x_i^2 \rangle = 2$ analytically, we compute $\langle y_i^2 \rangle$ by generating many exact samples. The posterior is shown in Figure \ref{fig:rosenbrock}.
    Table \ref{table} shows that for this example, the $q = 2$ Hamiltonian significantly underperforms relative to the variable
    mass choice. Compared to NUTS  
    the improvement is a factor of 4. 
    Comparing tuning-free version of 
    MCHMC to tuned version we find an 
    order of magnitude difference, indicating 
    that our automatic tuning procedure fails 
    for such extremely non-Gaussian 
    distributions. 
    MCHMC without bounces performs a factor 
    of 2-3 worse than MCHMC with bounces, 
    and ESH is another factor of 3 worse.

\subsection{Neal's funnel}
    
    \begin{figure}
        \makebox[\textwidth][c]{\includegraphics[scale = 0.27]{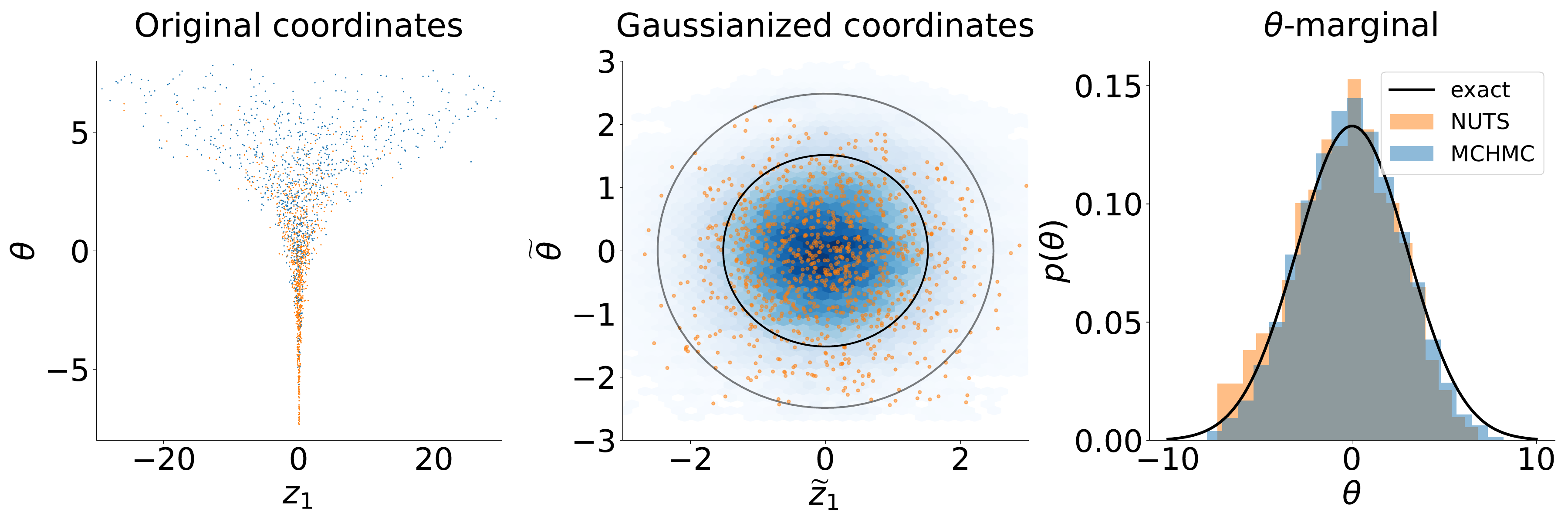}}
        \caption{The 20-dimensional Neal's funnel target.
        Left: the two-dimensional marginal in the ($\theta$, $z_1$) plane demonstrates the funnel shape of the target. 
        Middle: The two-dimensional marginal in the Gaussianized coordinates to exhibit better the quality of the 
        funnel posterior. We see that the samplers capture well the latent space even though they work in the hard-to-sample original space.
        Right: the one-dimensional marginal $\theta$ distribution.
        Both MCHMC and NUTS are capable of accurately sampling the funnel target, in contrast to results reported in \citet{ESH}. This
        highlights the importance of 
        bounces in MCHMC, and the importance
        of hyperparameter tuning in NUTS. 
        }
        \label{fig:funnel}
    \end{figure}
   
    This is a toy problem of a type that is typically encountered in the hierarchical Bayesian models \citep{HierarchicalWebsite}. The target density is \citep{NealHandbook}
    \begin{equation} \nonumber
        p(\theta, z_1,\, z_2,\, ... \, z_{d-1})= \mathcal{N}(\theta \vert 0, 3) \, \prod_{i = 1}^{d-1} \mathcal{N}(z_i \vert 0, e^{\theta / 2}),
    \end{equation}
    we take $d = 20$. We are interested in the posterior of the hyperparameter $\theta$. The problem is challenging for the HMC type of samplers because of the narrow funnel shape in which a large probability mass is hidden.
    The posterior is shown in Figure \ref{fig:funnel}. Table \ref{table} shows that for this example, the $q = 2$ Hamiltonian 
    drastically underperforms relative to variable mass MCHMC. Compared to NUTS (with warm-up) the improvement of MCHMC is a factor of 11. MCHMC without bounces and ESH fail to converge
    on this example.

\subsection{German credit}
    
    This is a popular Bayesian regression test case \citep{GermanCreditData}. We have real data about the costumers who applied for the credit at a bank, and we know the result of the approval process. We model the approval process as a Bayesian logistic model with 51 parameters and a sparsity inducing prior (see for example \citet{hoffmanGermanCredit}). The sampler is used to determine the posterior of the model parameters. We use the model implementation from the Inference Gym \citep{inferencegym} and initialize the sampler by a draw from a standard Gaussian, centered at the MAP solution. The ground truth moments are computed by a very long STAN run \citep{inferencegym}. We take the ESS for NUTS from \citet{DLMC}.
    Table \ref{table} shows that for this example, the $q = 2$ Hamiltonian also
    drastically underperforms relative to variable mass MCHMC. Compared to NUTS (with warm-up) the improvement of MCHMC is a factor of 13. MCHMC without bounces performs a factor 
    of 2-3 worse than with bounces, while ESH is considerably worse.

\subsection{Stochastic Volatility}
        
    \begin{figure}
        \makebox[\textwidth][c]{\includegraphics[scale = 0.3]{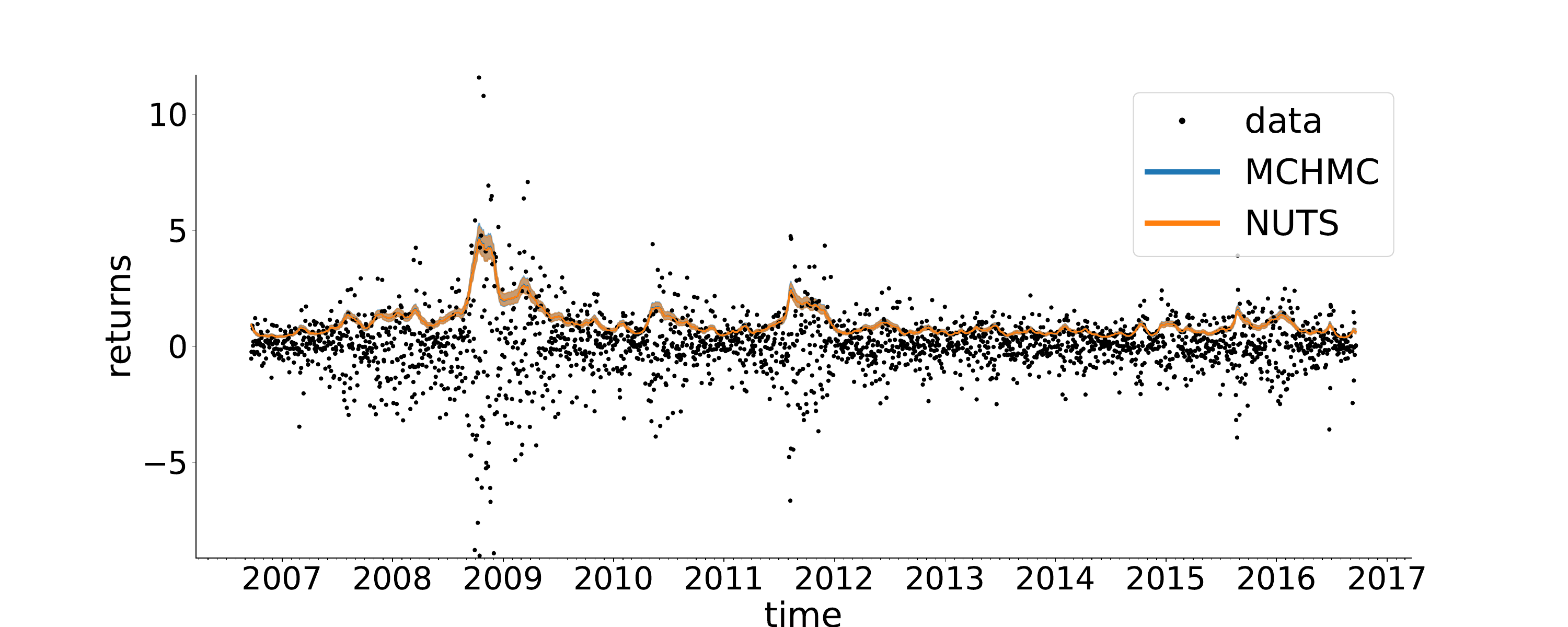}}
        \caption{The S\&P500 returns data are shown along with the volatility ($R_n$) posterior in the stochastic volatility model. We show the median and the first lower and upper quartiles. Both MCHMC and NUTS accurately capture the posterior.}
        \label{fig:stochastic volatility}
    \end{figure}
    
    This is a popular test case from the time series analysis \citep{NUTS, NummPyro}. We have $N = 2427$ values of the returns on the S\&P500 index $\{ r_n\}_{n = 1}^{N}$ in the time span of 10 years. The returns $r_n$ are modeled by a Student's-t distribution whose scale (volatility) $R_n$ is time varying and unknown. The prior for $\log R_n$ is a Gaussian random walk, with an exponential distribution of the random walk step-size $\sigma$. An exponential prior is also taken for the Student's-t degrees of freedom $\nu$. The generative process of the data is:
    \begin{align}
        &r_n / R_n \sim \text{Student's-t}(\nu) \qquad 
        &&\nu \sim \text{Exp}(\lambda = 1/10) \\ \nonumber
        &\log R_n \sim \mathcal{N}(\log R_{n-1}, \sigma) \qquad
        &&\sigma \sim \text{Exp}(\lambda = 1/0.02).
    \end{align}
    We use $\log \lambda \nu$ and $\log \lambda \sigma$ as parameters to make the configuration space unconstrained. The task is to find the posterior of the parameters $\{R_n\}_{n =1}^N$, $\sigma$ and $\nu$, given the observed data $\{r_n\}_{n =1}^N$.
    The ground truth moments are computed by a very long NUTS run. The posterior is shown in Figure \ref{fig:stochastic volatility}.
    Table \ref{table} shows that for this example, the standard $q=2$ Hamiltonian also
    drastically underperforms relative to $q=0$ MCHMC. Compared to NUTS (with warm-up) the improvement of MCHMC is a factor of 14-23. MCHMC without bounces and ESH fail to converge on the posterior.

\subsection{Cauchy distribution} \label{sec: cauchy}

Here, each parameter is standard Cauchy distributed:
\begin{equation}
    p(\x) = \prod_{i = 1}^d C(x_i) = \prod_{i = 1}^d \frac{1}{\pi} \frac{1}{1 + x_i^2}.
\end{equation}
We will use $d = 1000$. This is an example of a heavily tailed distribution. All moments are infinite, so we cannot define ESS through the error of the second moments as in the other examples. Instead, we will compute the squared bias of the entropy for each dimension and average over the dimensions:
\begin{equation}
    b_{\mathcal{L}}^2 = \frac{1}{d} \sum_{i = 1}^d \big( \mathbb{E}_{\text{sampler}} [-\log C(x_i)] - \mathbb{E}_{\text{truth}} [-\log C(x_i)] \big)^2
\end{equation}
The entropy of the Cauchy distribution is $\mathbb{E}_{\text{truth}} [-\log C(x_i)] = \log 4 \pi$.
In the limit of large number of dimensions, $b_{\mathcal{L}}^2$ converges to $\mathrm{Var}[- \log C] / n_{\mathrm{eff}}$, by the central limit theorem. $\mathrm{Var}[- \log C] = \pi^2 / 3$, so $b_{\mathcal{L}}^2 \approx 0.0165$ corresponds to $200$ effective samples. 

\begin{figure}
    \makebox[\textwidth][c]{\includegraphics[scale = 0.34]{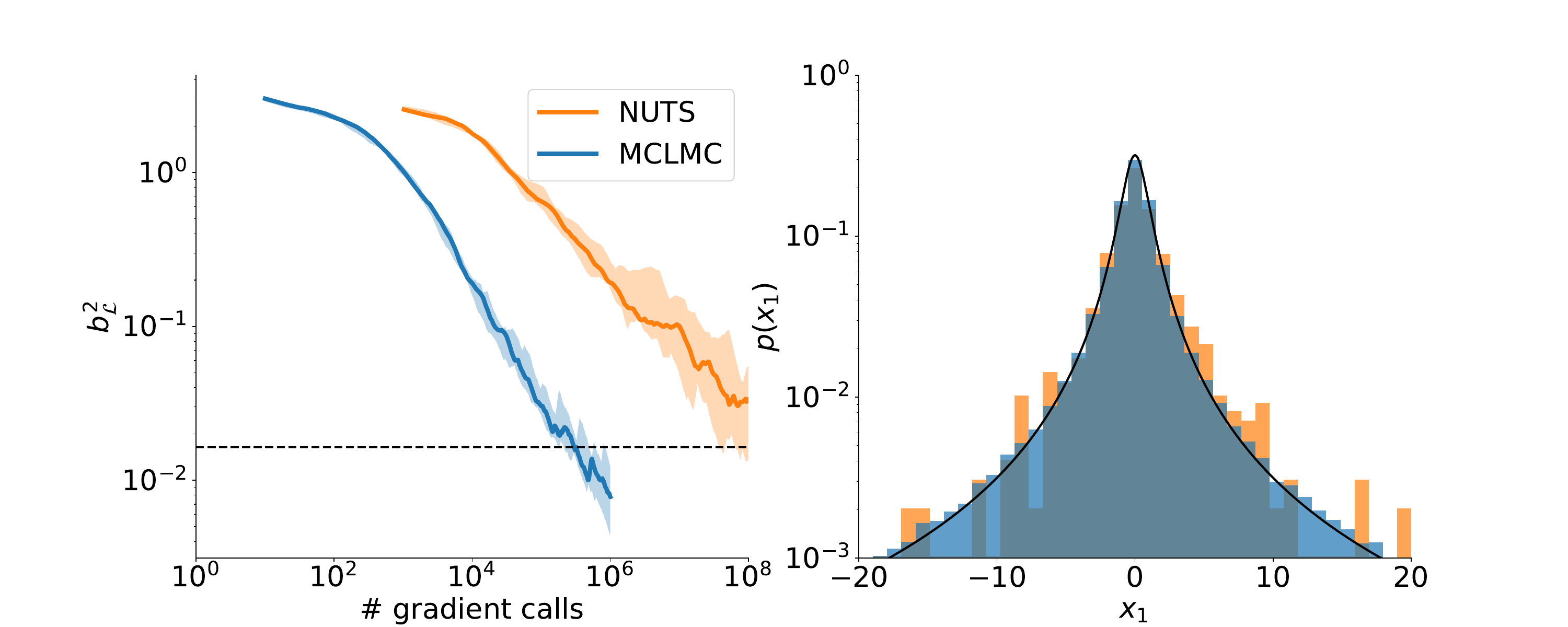}}
    \caption{1000 independent Cauchy distributed variables. 
    Left panel: the bias of the one dimensional entropy is shown as a function of gradient calls used. We ran 12 independent chains, the solid line corresponds to the median over the chains, the shaded region to 50\% of chains. The dotted line corresponds to $200$ effective samples. NUTS convergence is much slower than that of MCLMC. 
    Right panel: 1d posterior distribution for the $x_1$ parameter after using $10^6$ gradient calls. The ground truth Cauchy distribution is shown with a black line.}
    \label{fig: cauchy}
\end{figure}

We show the results in Figure \ref{fig: cauchy}. MCLMC was run for $10^6$ gradient calls, NUTS for $10^8$. Autotuning for MCLMC took additional $10^4$ samples and NUTS tuning took additional $10^6$ samples, but shorter tuning would also be possible. $L$ tuning of MCLMC is in principle problematic for this example, because it relies on the diverging variance of the parameters. However, at a relatively small sampling time of $10^4$ steps, the variances are still finite, and they give us some information about the typical scale of the distribution. 

The NUTS convergence is very slow, it only reaches $b_{\mathcal{L}}^2 = 0.03$ after $10^8$ gradient calls. The slow convergence in the tails is also apparent on the 1d marginal posterior plot in Figure \ref{fig: cauchy}.
MCLMC convergence is much faster, at $10^6$ calls it already produced more than $600$ effective samples. Its 1-d marginal distributions look nearly perfect.

\section{Discussion}

We have introduced Microcanonical Hamiltonian Monte Carlo (MCHMC) and 
Microcanonical Langevin-like Monte Carlo (MCLMC)
as a general class of energy conserving Hamiltonian dynamics models that can sample from the target 
distribution. One such Hamiltonian is recently introduced deterministic Energy Sampling Hamiltonian (ESH, \cite{ESH}), which 
however is not ergodic on its own. 
We introduce stochastic momentum decoherence via bounces as a general 
solution and an essential component 
of MCHMC. 
We propose occasional complete momentum decoherence (MCHMC) and 
continuous partial momentum decoherence (MCLMC) and found that MCLMC performs better on the ill-conditioned targets, but has a similar performance to MCHMC on other examples.
MCHMC and MCLMC  are stochastic algorithms, but 
take advantage of favorable 
deterministic dynamics of energy 
conserving Hamiltonians. As such 
the samples can be viewed as 
quasi-random in low dimensions, and indeed we 
achieve $ESS>1$ in very low 
dimensions. 

We developed an algorithm for tuning of 
the hyperparameters, which we found to be 
close to optimal over a wide range 
of targets. This further improves 
the wall-clock performance, since tuning of samplers often requires significant 
computational cost. MCHMC and MCLMC only have two 
 hyperparameters: the rate of momentum decoherence (or alternatively the bounce frequency) and the step size for the dynamics integrator. Our proposed tuning-free algorithm relates the decay-distance of the momentum correlations to the typical set size.  Microcanonical nature of MCHMC 
implies that standard Metropolis 
Adjustment is not possible, and 
instead MCHMC bias must be controlled 
by the choice of step size to make it
smaller than variance. We relate 
the bias to the error in 
 energy, and for sufficiently small
 energy error the bias
 is also small. MCHMC is less 
 sensitive than unadjusted HMC to catastrophic 
 integration errors which lead
 the dynamics to high target $\mathcal{L}$ above the typical set: such samples are 
 downweighted by the weights $w$ in MCHMC. 
 This is because the dynamics of $q=0$
 MCHMC is very different from HMC: while HMC moves fastest when 
 target $\mathcal{L}$ is low, MCHMC moves fastest when $\mathcal{L}$ is high. 



In Appendix \ref{sec: geodesic} we 
provide arguments for ergodicity 
of bounce-based momentum decoherence. We show that the Hamiltonian dynamics of the proposed Hamiltonians is equivalent to the geodesic motion on a conformally flat manifold, with conformal factor proportional to the target distribution. In particular, for $q=0$ MCHMC, the target distribution is identified with the Riemannian volume form. Due to the bounces, our algorithm is then performing an approximated geodesic random walk on this manifold in the sense of \cite{Jrgensen1975TheCL}. The latter is known to be ergodic \citep{sunada1983mean}, and it may be 
possible to prove ergodicity of our approximated walk as well.
This can be supplemented by other ergodic strategies, such as boundary reflections enhancing chaos \citep{PhysRevLett.77.2941}. Finally, 
while the class of Hamiltonian 
models is already large due to the 
many possible choices of the specific 
form of Hamiltonian, 
the geometric picture developed in Appendix \ref{sec: geodesic} 
further generalizes it via  
an even larger 
class of models that can sample
from the target distribution.



\acks{This material is based upon work supported in part by the Heising-Simons Foundation grant 2021-3282 and by the U.S. Department of Energy, Office of Science, Office of Advanced Scientific Computing Research under Contract No. DE-AC02-05CH11231 at Lawrence Berkeley National Laboratory to enable research for Data-intensive Machine Learning and Analysis. The work of ES and GBDL is supported in part by the Simons Foundation, by the National Science Foundation grant number PHY-1720397, and  the Stanford Research Computing Center.  We thank Qijia Jiang for useful discussions.}

\appendix

\section{Geodesic motion on a curved geometry} \label{sec: geodesic}
In this Appendix, we provide a geometric picture of the Hamiltonian approach
to sampling we developed in the main text. Specifically, we review how at
fixed energy the Hamiltonian trajectories in configuration space can be mapped
to geodesics of an appropriate metric defined in the same space, known as
\textit{Jacobi metric}. This approach connects the ergodic and chaotic
theories for dynamical system to the geometric ones, providing a
complementary intuition for the dynamics. 

To fix the notation, we use local coordinates $x^i$, \ $i = 1, \ldots, d$ ,
on the $d$-dimensional configuration space $X$ and denote by $d  s^2 =
g_{i j} (\mathbf{x}) \text{dx}^i \text{dx}^j$ a Riemannian metric on it;
Einstein's sum notation for repeated indices is assumed throughout. A curve
$\mathbf{x} (\sigma)$ between any two points $\mathbf{x_1}$ and
$\mathbf{x_2}$ is said to be a \emph{geodesic} if it locally minimizes the
Riemannian distance between them (i.e. any small deformation of $\mathbf{x}(\sigma)$ increases the
total length of the curve). In standard parametrization, it satisfies in local
coordinates the geodesic equation
\begin{equation}
  \frac{d}{d \sigma} \left( g_{i j}  \frac{d x^j}{d \sigma} \right) =
  \frac{1}{2} \frac{\partial g_{k l}}{\partial x^i} \frac{d x^k}{d \sigma}
  \frac{d x^l}{d \sigma} \hspace{4em} i = 1, \ldots d, \label{eq:geoEq}
\end{equation}
which is a system of $d$ second order ordinary differential equations\footnote{
We are always free to reparametrize the coordinate $\sigma$ along the curve.
This will change the form of the differential equation \eqref{eq:geoEq}, but
not the shape of the curve on $X$.
}.

In the conformally flat case $g_{i j} (\mathbf{x}) = e^{f
(\mathbf{x})} \delta_{i j}$, with $\delta_{i j}$ being the Kronecker delta, Equation \eqref{eq:geoEq} simplifies to
\begin{equation}
  \frac{d}{d \sigma} \left( e^f  \frac{d x^i}{d \sigma} \right) = \frac{1}{2}
  \kappa^2 \frac{\partial f}{\partial x^i} \,, \label{eq:confGeo}
\end{equation}
where $\kappa^2 \equiv e^f \frac{d x^i}{d \sigma} \frac{d x^j}{d \sigma}
\delta_{i j}$ is a constant of motion. A direct connection to the Hamiltonian
formalism we developed in the main text is obtained by noticing that the
continuum Hamilton-Jacobi Equations \eqref{eq: hamilton eqs} for the variable mass Hamiltonian \eqref{eq: H1} on a given energy hypersurface $E$ coincide with Equation
\eqref{eq:confGeo} upon the identification
\begin{equation}
  e^{f (\mathbf{x})} = m (\mathbf{x}), \quad \kappa^2 = 2 E ,\quad t = \sigma.
\end{equation}
in terms of the target density distribution  $m
(\mathbf{x}) = p (\mathbf{x})^{2 / d}$ and the corresponding Jacobi metric
reads
\begin{equation}\label{eq:metRL}
  g_{i j} (\mathbf{x}) \equiv p (\mathbf{x})^{2 / d} \delta_{i j},
  \hspace{4em} \text{ for variable mass Hamiltonian.} 
\end{equation}
Equation \eqref{eq:metRL} provides a very natural geometric interpretation of
the target distribution as the Riemannian volume density. Indeed, the latter
is computed as the square root of the metric determinant, which in terms
of the target density gives
\begin{equation}
  \sqrt{\det (g (\mathbf{x}))} = p (\mathbf{x}) .
\end{equation}
Summarizing, trajectories $\mathbf{x} (t)$ of the variable mass Hamiltonian \eqref{eq: H1} at a
fixed energy $E$, with $m (\mathbf{x}) = p (\mathbf{x})^{2 / d}$, are
geodesics for the metric (Equation \eqref{eq:metRL}) which solve the geodesic equation
parametrized as in Equation \eqref{eq:confGeo}, with $t = \sigma$. For any function
$\mathcal{O} : X \rightarrow \mathbb{R}$, \ then ergodicity in the Hamiltonian
sense implies (upon marginalization over the momenta, as in Equation \eqref{eq: marginal condition})
\begin{equation}
  \langle \mathcal{O} \rangle \equiv \int_X p (\mathbf{x})  \mathcal{O}
  (\mathbf{x}) d \mathbf{x} \propto \frac{1}{T} \int_0^T d t \mathcal{O}
  (\mathbf{x} (t))
\end{equation}
where $\mathbf{x} (t)$ solves the Hamilton-Jacobi equation for Equation \eqref{eq: H1}. From the
discussion above,  this is equivalent to ergodicity along Riemannian
geodesics, i.e. the notion that geodesic distribute accordingly to the Riemannian
volume form. Indeed,
\begin{equation}
  \langle \mathcal{O} \rangle = \langle \mathcal{O} \rangle_g \equiv \int_X
  \sqrt{g}  \mathcal{O} (\mathbf{x}) d \mathbf{x} \propto \frac{1}{T}
  \int_0^T d \sigma \mathcal{O} (\mathbf{x} (\sigma))
\end{equation}
when $\mathbf{x} (\sigma)$ solves the geodesic Equation \eqref{eq:confGeo}.

\

Thanks to this observation, we can use the geometric description as another way to gain intuition about the trajectories and their ergodicity/mixing properties.
The idea of studying the properties of Hamiltonian systems through their geometric description is not new, see for example \citet{pettini2007geometry, di2021hamiltonian} for reviews of chaos in Hamiltonian systems from the geometric point of view, and \citet{seiler2014positive} for a direct application of the geometric methods to the analysis of standard Hamiltonian Monte Carlo samplers.



In particular, curvatures control the behavior of the geodesics and their spreading, and thus the distribution of the trajectories in configuration space. More precisely, take a point $\mathbf{x}\in X$ and two geodesics $\gamma_1(\sigma)$ and $\gamma_2(\sigma)$ that at $\sigma = 0$  pass through $\mathbf{x}$ and are directed along two directions $\xi_1$ and $\xi_2$, respectively. The \emph{sectional curvature} along the $\xi_1$-$\xi_2$ plane, $K(\xi_1,\xi_2)$, directly controls the distance between the two geodesics as they evolve:
\begin{equation}
    \text{dist}(\gamma_1(\sigma), \gamma_2(\sigma)) = \sqrt{2}\sigma \left(1-\frac{1}{12}K(\xi_1,\xi_2) \sigma^2+O(\sigma^3)\right)\qquad \qquad \sigma\to 0\quad.
\end{equation}
Negative sectional curvatures ($K<0$) tend to spread the geodesics.
On the conformally flat manifold relevant for our discussion, the sectional curvature along the $i$-$j$ plane reads 
\begin{equation}
    K_{ij} = -F^{-2}\left(\partial_i^2 f + \partial_j^2 f +\sum_{k\ne i,j}(\partial_k f)^2\right)
\end{equation}
with $f=\log(F)$. We find that, for example, for a Gaussian distribution of unit width, $m\propto \exp(-{\bf x}^2/d)$, the sectional curvatures are positive near the peak but almost all negative outside the typical set ${\bf x}^2\sim d$.  We note also that it is a finite proper distance to $|{\bf x}|=\infty$, with the proper area of that region approaching zero while the magnitude of the negative curvatures blows up. As in the Hamiltonian formulation, a boundary condition is required to complete the specification of the system.  The ESH algorithm \citep{ESH} prescribes steps of vanishing proper distance in the $|\x|\to\infty$ limit.

The idea of diagnosing chaos and mixing properties of dynamical systems through negative curvature goes back to \cite{krylov} and mathematical proofs of ergodicity and chaos in the Riemannian settings exist for the case of purely negative curvature (see \cite{valva2019manifolds} for a recent review.) 
Even though negative curvature does not seem to be strictly needed for chaos, since also sufficiently varying positive curvature can enhance the mixing properties (e.g. \cite{pettini2007geometry, seiler2014positive}), in our algorithm bounces play the role of ensuring ergodicity where needed, such as the positively curved regions near Gaussian peaks where the curvature might not be enough.  

\subsection{Other Hamiltonians in the geometric framework}

So far, we connected the variable mass model to the geodesic motion picture. 
More broadly, we can ask whether also the more general separable Hamiltonian \eqref{separable}, \eqref{eq:qcases}, with $q>0$ and the
relativistic Hamiltonian \eqref{eq: H2}, and the corresponding samplers, can be understood as geodesic motion.
The answer turns out to be affirmative, with the $q = 2$ already discussed by Jacobi (see e.g. the review in \citet{pettini2007geometry}) and the relativistic case analyzed in \citet{gibbons2015jacobi}. 
In particular, for these more general cases, the identification also requires a non-trivial identification
between the geodesic time $\sigma$ and the Hamiltonian time $t$.
For a generic separable Hamiltonian \eqref{separable}, \eqref{eq:qcases} with $q > 0$ this results in
\begin{equation}
  H = \frac{1}{q} \mathbf{|\Pi|}^q + V (\mathbf{x}) \qquad \rightarrow
  \qquad g_{i  j} (\mathbf{x}) = (E - V (\mathbf{x}))^\frac{2}{q} \delta_{i
   j} \;, \quad \frac{d \sigma}{\text{dt}} =q^{\frac{2}{q}-1} (E - V (\mathbf{x}))\:,
\end{equation}
while for the relativistic BI Hamiltonian,  \eqref{eq: H2} with $V(\mathbf{x}) = c^2(\mathbf{x})$, we have
\begin{equation}
  H = \sqrt{V (\mathbf{x}) (V (\mathbf{x}) + \mathbf{\Pi}^2)} \qquad
  \rightarrow \qquad g_{i  j} (\mathbf{x}) = \frac{E^2 - V
  (\mathbf{x})^2}{V (\mathbf{x})} \delta_{i  j} \;, \quad \frac{d
  \sigma}{\text{dt}} = E^2 - V (\mathbf{x})^2\:.
\end{equation}
In these cases, upon identification of $V (\mathbf{x})$ with $p
(\mathbf{x})$ through \eqref{eq: tuning3} and \eqref{eq: tuning2} respectively, the volume density is not simply proportional
to $p (\mathbf{x})$. This is taken into account by the fact that steps
in Hamiltonian time $t$ need to be weighted with the non-trivial $t (\sigma)$
to map them into steps in geodesic ``time'' $\sigma$:
\begin{eqnarray}
  \int_X p (\mathbf{x})  \mathcal{O} (\mathbf{x}) d \mathbf{x} & \propto
  & \frac{1}{T} \int_0^T d t \mathcal{O} (\mathbf{x} (t)) \\
  & = & \frac{1}{\sigma (t = T)} \int_{\sigma (t = 0)}^{\sigma (t = T)} d
  \sigma \frac{d t}{d \sigma}  \mathcal{O} (\mathbf{x} (\sigma)) \\
  & \propto & \int_X \sqrt{g}  \frac{d t}{d \sigma}  \mathcal{O}
  (\mathbf{x}) d \mathbf{x} \,.
\end{eqnarray}

Similar results can be obtained for more general Hamiltonians with speed limits enforced by different mechanisms, such as logarithmic branch cuts, as in \cite{Mathis:2020vdn}. We leave a more comprehensive studies of the general possibility for future work.

Transforming back, we can collect the results of this Appendix, regarding the examples of microcanonical sampling studied in the present work, in the
following
\begin{proposition}\label{prop: geo}
 Microcanonical sampling (with isotropic kinetic term) from a target
 distribution $p (\mathbf{x}) $, $\mathbf{x} \in \mathbb{R}^d$, is equivalent to geodesic
 evolution on the Riemannian manifold $\left(\mathbb{R}^d, g\right)$, where $g$ is the conformally flat
  metric on $\mathbb{R}^d$
  \begin{equation}
    d  s^2 = (\gamma (\mathbf{x}) p (\mathbf{x}))^{2 / d} \delta_{i j} d x^i d x^j
    \quad, \label{eq:metgamma}
  \end{equation}
  and $\gamma (\mathbf{x})$ is a positive real function.  Indeed, if the system is ergodic,
  expectation values of general functions $\mathcal{O}: \mathbb{R}^d \to \mathbb R$ can be computed as
  \begin{equation}
    \langle \mathcal{O} \rangle_p = \int p (\mathbf{x}) \mathcal{O} (\mathbf{x}) d^d \mathbf{x} \propto
    \frac{1}{\Sigma} \int_0^{\Sigma} d \sigma \gamma^{- 1} (\mathbf{x} (\sigma))
    \mathcal{O} (\mathbf{x} (\sigma)) \label{eq:rhs}
  \end{equation}
  where $\mathbf{x} (\sigma)$ solves the following geodesic equation for
  \eqref{eq:metgamma}
  \begin{equation}\label{eq:geoEqProp}
    \frac{d}{d \sigma} \left(e^{f} \frac{d x^i}{d \sigma}\right) = \frac{1}{2} \frac{\partial f}{\partial x^i}
    \hspace{4em} e^{f} = (\gamma (\mathbf{x}) p (\mathbf{x}))^{2 / d} \qquad\qquad i = 1,\dots,d
    \, .
  \end{equation}
  Specifically, for the cases analyzed in this paper:
  \begin{enumerate}
    \item Variable mass \eqref{eq: H1}: $\gamma(\mathbf{x}) = E$ = constant.
    
    \item Relativistic BI Hamiltonian \eqref{eq: H2}: $\gamma(\mathbf{x}) = E^2 - c(\mathbf{x})^4$
    where the relation between $c(\mathbf{x})$ and $p(\mathbf{x})$ is given by \eqref{eq: tuning2}.
    
      \item Separable Hamiltonian \eqref{separable}, \eqref{eq:qcases}, with $q > 0$ : \ $\gamma(\mathbf{x}) = p (\mathbf{x})^{\frac{q}{d - q}}$
  \end{enumerate}
\end{proposition}

Notice that one can remove the weight factor $\gamma^{- 1}$ in \eqref{eq:rhs} using the freedom to redefine the time coordinate along the trajectory. This has the effect of modifying the form of the equation \eqref{eq:geoEqProp} with an extra $\gamma$-dependence:
\begin{equation}
\frac{d}{d \tau}\left(\gamma^{-1} e^f \frac{d x^i}{d\tau}\right) = \frac{1}{2}\gamma \frac{\partial f}{\partial x^i}    
\end{equation}
where the new time coordinate $\tau $ is defined by $\frac{d \sigma}{d\tau} = \gamma$.
For example, specializing to the $q$-Hamiltonians
\eqref{eq:qcases} results in the geodesics equation
\begin{equation}\label{eq:geoUnw}
\frac{d}{d \tau}\left(p^\frac{q-2}{q-d}\frac{d x^i}{d\tau}\right) = \frac{1}{q}\frac{\partial \left(p^\frac{q}{d-q}\right)}{\partial x^i} \,.
\end{equation}
Fixing $q >0 \, (\neq  d)$ any integration scheme for  the geodesics equation \eqref{eq:geoUnw} produces a microcanonical unweighted sampler. For $q = 0$ already the integration of the original geodesic equation \eqref{eq:geoEqProp} produced an unweighted sampler.

The analysis in this section has been performed for the isotropic cases studied in this paper, but it can straightforwardly be
extended to the more general anistropic case, in which case the Jacobi metric is not conformally flat.
Moreover, this rewriting suggests that choices for $\gamma(\mathbf{x})$ different from the ones we analyzed in this work can produce new samplers, whose performance could be interesting to explore.

\vskip 0.2in

\bibliography{citations}

\end{document}